\documentclass[12pt]{article}
\usepackage{mathrsfs}
\usepackage[T1]{fontenc}
\usepackage{mathpazo}
\usepackage{setspace}
\usepackage{cite}
\usepackage{amsfonts}
\usepackage{amssymb}
\usepackage{amsmath}
\usepackage{esint}                   % For closed integrals
\usepackage{epsfig}
\usepackage{latexsym}
\usepackage{color}
\usepackage{graphicx}
\usepackage{hyperref}
\usepackage{caption}
\usepackage{nicefrac}
\usepackage[latin1]{inputenc}
\usepackage{pstricks}
\usepackage{slashed}
\usepackage{multirow}
\def\hybrid{\topmargin -30pt    \oddsidemargin 0pt %%%%%%%%%%%%%% Archive-30pt
        \headheight 0pt \headsep 0pt
        \textwidth 6.25in       % A4 paper
        \textheight 9.5in       % A4 paper
        \marginparwidth .875in
        \parskip 5pt plus 1pt   \jot = 1.5ex}

\hybrid

\def\baselinestretch{1.2}

\catcode`\@=11

\def\marginnote#1{}
\newcount\hour
\newcount\minute
\newtoks\amorpm
\hour=\time\divide\hour by60
\minute=\time{\multiply\hour by60 \global\advance\minute by-\hour}
\edef\standardtime{{\ifnum\hour<12 \global\amorpm={am}%
        \else\global\amorpm={pm}\advance\hour by-12 \fi
        \ifnum\hour=0 \hour=12 \fi
        \number\hour:\ifnum\minute<10 0\fi\number\minute\the\amorpm}}
\edef\militarytime{\number\hour:\ifnum\minute<10 0\fi\number\minute}
\def\draftlabel#1{{\@bsphack\if@filesw {\let\thepage\relax
   \xdef\@gtempa{\write\@auxout{\string
      \newlabel{#1}{{\@currentlabel}{\thepage}}}}}\@gtempa
   \if@nobreak \ifvmode\nobreak\fi\fi\fi\@esphack}
        \gdef\@eqnlabel{#1}}
\def\@eqnlabel{}
\def\@vacuum{}
\def\draftmarginnote#1{\marginpar{\raggedright\scriptsize\tt#1}}

\def\draft{\oddsidemargin -.5truein
        \def\@oddfoot{\sl preliminary draft \hfil
        \rm\thepage\hfil\sl\today\quad\militarytime}
        \let\@evenfoot\@oddfoot \overfullrule 3pt
        \let\label=\draftlabel
        \let\marginnote=\draftmarginnote
   \def\@eqnnum{(\theequation)\rlap{\kern\marginparsep\tt\@eqnlabel}%
\global\let\@eqnlabel\@vacuum}  }

\def\draft2{
        \def\@oddfoot{\sl preliminary draft \hfil
        \rm\thepage\hfil\sl\today\quad\militarytime}
        \let\@evenfoot\@oddfoot \overfullrule 3pt
        \let\label=\draftlabel
        \let\marginnote=\draftmarginnote
   \def\@eqnnum{(\theequation)\rlap{\kern\marginparsep\tt\@eqnlabel}%
\global\let\@eqnlabel\@vacuum}  }

\def\preprint{\twocolumn\sloppy\flushbottom\parindent 2em
        \leftmargini 2em\leftmarginv .5em\leftmarginvi .5em
        \oddsidemargin -.5in    \evensidemargin -.5in
        \columnsep .4in \footheight 0pt
        \textwidth 10.in        \topmargin  -.4in
        \headheight 12pt \topskip .4in
        \textheight 6.9in \footskip 0pt
        \def\@oddhead{\thepage\hfil\addtocounter{page}{1}\thepage}
        \let\@evenhead\@oddhead \def\@oddfoot{} \def\@evenfoot{} }

\def\numberbysection{\@addtoreset{equation}{section}
        \def\theequation{\thesection.\arabic{equation}}}

\def\underline#1{\relax\ifmmode\@@underline#1\else
        $\@@underline{\hbox{#1}}$\relax\fi}

\def\titlepage{\@restonecolfalse\if@twocolumn\@restonecoltrue\onecolumn
     \else \newpage \fi \thispagestyle{empty}\c@page\z@
        \def\thefootnote{\fnsymbol{footnote}} }

\def\endtitlepage{\if@restonecol\twocolumn \else \newpage \fi
        \def\thefootnote{\arabic{footnote}}
        \setcounter{footnote}{0}}  %\c@footnote\z@ }

\catcode`@=12
\relax

\def\figcap{\section*{Figure Captions\markboth
        {FIGURECAPTIONS}{FIGURECAPTIONS}}\list
        {Figure \arabic{enumi}:\hfill}{\settowidth\labelwidth{Figure
999:}
        \leftmargin\labelwidth
        \advance\leftmargin\labelsep\usecounter{enumi}}}
 \relax
\def\tablecap{\section*{Table Captions\markboth
        {TABLECAPTIONS}{TABLECAPTIONS}}\list
        {Table \arabic{enumi}:\hfill}{\settowidth\labelwidth{Table
999:}
        \leftmargin\labelwidth
        \advance\leftmargin\labelsep\usecounter{enumi}}}
 \relax
\def\reflist{\section*{References\markboth
        {REFLIST}{REFLIST}}\list
        {[\arabic{enumi}]\hfill}{\settowidth\labelwidth{[999]}
        \leftmargin\labelwidth
        \advance\leftmargin\labelsep\usecounter{enumi}}}
 \relax
\makeatletter
\newcounter{pubctr}
\def\publist{\@ifnextchar[{\@publist}{\@@publist}}
\def\@publist[#1]{\list
        {[\arabic{pubctr}]\hfill}{\settowidth\labelwidth{[999]}
        \leftmargin\labelwidth
        \advance\leftmargin\labelsep
        \@nmbrlisttrue\def\@listctr{pubctr}
        \setcounter{pubctr}{#1}\addtocounter{pubctr}{-1}}}
\def\@@publist{\list
        {[\arabic{pubctr}]\hfill}{\settowidth\labelwidth{[999]}
        \leftmargin\labelwidth
        \advance\leftmargin\labelsep
        \@nmbrlisttrue\def\@listctr{pubctr}}}
 \relax
\makeatother

\def\be{\begin{equation}}
\def\ee{\end{equation}}
\def\ba{\begin{eqnarray}}
\def\ea{\end{eqnarray}}

\def\b{\beta}

\def\d{\delta}

\def\m{\mu}
\def\n{\nu}
\def\om{\omega}

\def\l{\lambda}

\def\s{\sigma}

\def\cA{{\cal A}}
\def\cB{{\cal B}}
\def\cD{{\cal D}}
\def\cE{{\cal E}}

\def\cM{{\cal M}}

\def\cT{{\cal T}}

\def\no{\noindent}

\def\IR{\relax{\rm I\kern-.18em R}}

\def\inv{^{\raise.0ex\hbox{${\scriptscriptstyle -}$}\kern-.05em 1}}

\def \ov {\over}

\def\diag{{\rm diag}}

\def\fe{\mathfrak{e}}
\def\hfe{\hat{\mathfrak{e}}}

\begin{document}
%\draft2

%\renewcommand{\theequation}{\arabic{equation}}
%\renewcommand{\theequation}{\thesection.\arabic{equation}}

\renewcommand{\theequation}{\thesection.\arabic{equation}}
\csname @addtoreset\endcsname{equation}{section}

\begin{titlepage}
\begin{center}

\hfill HU-EP-24/36

\phantom{xx}
\vskip 0.5in

{\large \bf Type-II backgrounds from deformed coset CFTs}

\vskip 0.5in

{\bf Georgios Itsios}${}^{1a}$\phantom{x}  % and \phantom{x} {\bf ...}${}^{2b}$ \vskip 0.1in

${}^1$ Institut f\"{u}r Physik, Humboldt-Universit\"{a}t zu Berlin,\\
IRIS Geb\"{a}ude, Zum Gro{\ss}en Windkanal 2, 12489 Berlin, Germany\\

\vskip .2in

%${}^2$  Afiliation of the co-authors

\end{center}

\vskip .4in

\centerline{\bf Abstract}

\no We construct a plethora of type-II supergravity solutions featuring AdS factors in their geometries, derived from integrable deformations of coset CFTs. Specifically, we uplift the $\l$-deformed models of $\nicefrac{SO(4)_k}{SO(3)_k}$ and $\nicefrac{SO(5)_k}{SO(4)_k}$, including the non-compact version of the latter. Requiring reality for the backgrounds imposes bounds on the deformation parameter. As a result, some of the solutions do not admit non-Abelian T-dual limit.

\vfill
\no
 {%\footnotesize
$^a$georgios.itsios@physik.hu-berlin.de  
%$^b$e-mail of the co-authors
}

\end{titlepage}
\vfill
\eject

%\def\baselinestretch{1.2}
%\baselineskip 10 pt
%\noindent

%\tableofcontents

\def\baselinestretch{1.2}
\baselineskip 20 pt

\newcommand{\eqn}[1]{(\ref{#1})}

\tableofcontents

%%%%%%%%%%%%%%%%%%%%%%%%%%%%%%%%%%%%%%%%%%%%%%%%%%%%%%%%%%%%%%
\section{Introduction}

Integrability preserving deformations of two-dimensional $\s$-models garnered considerable attention in the literature. A notable example is the so-called $\l$-model \cite{Sfetsos:2013wia}, originally formulated by gauging the combined action of a Wess-Zumino-Witten (WZW) model and a principal chiral model (PCM) for a semi-simple group $G$. This model interpolates between an exact conformal field theory (CFT), represented by a WZW model with a current algebra symmetry at level $k$, and the non-Abelian T-dual (NATD) of the PCM for $G$. This construction can be generalised to deformations based on cosets with Lagrangian description through gauged WZW models, as well as symmetric or semi-symmetric spaces \cite{Sfetsos:2013wia,Hollowood:2014rla,Hollowood:2014qma}. A variation of the $\l$-model with spectators was derived in \cite{Borsato:2024alk} by modifying the gauging procedure. Further generalisations include \cite{Georgiou:2016zyo,Georgiou:2017jfi,Georgiou:2018hpd,Georgiou:2018gpe,Driezen:2019ykp,Sfetsos:2017sep}. Another class of integrable $\s$-models connected to the PCM on groups, known as $\eta$-deformations, was introduced in \cite{Klimcik:2002zj,Klimcik:2008eq}. This was generalised to the context of coset spaces in \cite{Delduc:2013fga,Delduc:2013qra}, inspiring a lot of activity in the framework of the AdS/CFT correspondence. A relation between the isotropic $\l$-model and the $\eta$-model, by means of Poisson-Lie T-duality \cite{Klimcik:1995ux} and appropriate analytic continuations, has been established in \cite{Vicedo:2015pna,Hoare:2015gda,Sfetsos:2015nya}.

The main objective of the present work is to build solutions of type-II supergravity out of the background fields of $\l$-deformed cosets. This is in general a non-trivial task, which requires the completion of the $\s$-model field content with appropriate Neveu-Schwarz (NS) and Ramond-Ramond (RR) fields. Nevertheless, type-II supergravity backgrounds based on $\l$- and $\eta$-deformed (super) cosets have been constructed in various works \cite{Hoare:2015gda,Demulder:2015lva,Borsato:2016zcf,Itsios:2019izt,Chervonyi:2016ajp,Borsato:2016ose,Hoare:2015wia,Hoare:2018ngg,Lunin:2014tsa,Hoare:2022asa}. More recently, the $\l$-models on the groups $SL(2, \mathbb{R})$ and $SU(2)$ have been promoted to type-II supergravity backgrounds
\footnote{
In fact, this was first achieved in \cite{Sfetsos:2014cea}. However, the RR fluxes of the supergravity solution found there are imaginary. The background can be made real by an appropriate analytic continuation in the $SL(2, \mathbb{R})$ coordinates \cite{Itsios:2023kma}. This corresponds to a different patch where the $SL(2, \mathbb{R})$ group element is not connected to the identity.
}
\cite{Itsios:2023kma,Itsios:2023uae}, describing deformations of the near-horizon limit of NS1 \& NS5 brane intersections. Following the approach of \cite{Itsios:2019izt}, we lift the $\l$-models on $\nicefrac{SO(4)_k}{SO(3)_k}$, $\nicefrac{SO(5)_k}{SO(4)_k}$ and $\nicefrac{SO(1,4)_{-k}}{SO(4)_{-k}}$ to ten dimensions, such that the type-II supergravity equations of motion are satisfied. Our motivation is driven by the ideas of holography and therefore we focus on backgrounds that exhibit AdS factors in their geometries. Our findings are summarised in Tables \ref{ListOfSolutionsCS3lambda} \& \ref{ListOfSolutionsCS4lambda}.

The rest of the paper is organised as follows: In Section \ref{Section2}, we review the background field content of the $\l$-models on $\nicefrac{SO(4)_k}{SO(3)_k}$, $\nicefrac{SO(5)_k}{SO(4)_k}$ as well as their non-compact versions. Section \ref{Section3} is devoted to the embedding of the $\l$-model on $\nicefrac{SO(4)_k}{SO(3)_k}$ into type-IIA and type-IIB supergravities. In Section \ref{Section4}, we discuss embeddings of the $\nicefrac{SO(5)_k}{SO(4)_k}$ and $\nicefrac{SO(1,4)_{-k}}{SO(4)}_{-k}$ $\l$-models. Conclusions and future ideas are contained in Section \ref{Section5}. We have also included an appendix, where we review the field content and the equations of motion of the type-IIA and type-IIB supergravities.

%%%%%%%%%%%%%%%%%%%%%%%%%%%%%%%%%%%%%%%%%%%%%%%%%%%%%%%%%%%%%%
\section{Review of the $\l$-deformed $\s$-models}
\label{Section2}

In this section, we summarise the background fields of the $\l$-models on $\nicefrac{SO(4)_k}{SO(3)_k}$ and $\nicefrac{SO(5)_k}{SO(4)_k}$ \cite{Demulder:2015lva}. We also discuss the content of their non-compact counterparts, which are obtained by appropriate analytic continuations.

%%%%%%%%%%%%%%%%%%%%%%%%%%%%%%%%%%%%%%%%%%%%%%%%%%%%%%%%%%%%%%
\subsection{The $\nicefrac{SO(4)_k}{SO(3)_k}$ $\l$-deformed model}
\label{CS3Lambda}

In the case of the $\nicefrac{SO(4)_k}{SO(3)_k}$ $\l$-deformed coset, the background fields are limited to the target space metric and a scalar. The metric can be conveniently expressed using the frame
\begin{equation}
 \label{CS3lambdaFrame}
 \begin{aligned}
  & \fe^1 = \frac{2 \, \l_-}{\sqrt{\cA}} \Big( - \om \, x \, dx + \frac{y}{\om} \, dy + \frac{\cA}{\om^2_+} \, d\om \Big) \, ,
  \\[5pt]
  & \fe^2 = - \frac{2 \, \l_+}{\sqrt{\cA \, \cD}} \Big( \cD \, \om \, dx + \frac{x \, y}{\om} \, dy + \frac{\cA \, x}{\om^2_+} \, d\om \Big) \, ,
  \\[5pt]
  & \fe^3 = \frac{2 \, \l_+}{\sqrt{\cD}} \Big( \frac{dy}{\om} - \frac{y}{\om^2_+} \, d\om \Big) \, .
 \end{aligned}
\end{equation}
The signature of the three-dimensional space is $(+ , + , +)$. To simplify the expressions, we define the following functions
\begin{equation}
 \label{OmegaPlusMinusAD}
 \om_\pm = \sqrt{1 \pm \om^2} \, , \qquad \cA = 1 - x^2 - y^2 \, , \qquad \cD = 1 - x^2
\end{equation}
and the constants
\begin{equation}
 \label{LambdaPlusMinus}
 \l_\pm = \sqrt{k \frac{1 \pm \l}{1 \mp \l}} \, , \qquad 0 \leq \l < 1 \, .
\end{equation}
Along with the metric, there is a scalar field $\Phi$ that depends on the coordinates $(x , y , \om)$ through the following combination
\begin{equation}
 \label{CS3lambdaScalar}
 e^{- 2 \Phi} = 8 \, \frac{\cA \, \om^2}{\om^4_+} \, .
\end{equation}
From here on, this deformed coset CFT will be referred as $CS^3_{\l}$.

One can verify that the metric and the scalar of the $CS^3_{\l}$ model satisfy
\begin{equation}
 \label{CS3lambdaBetaFunctions}
 \begin{aligned}
  & \b^{\Phi} = R + 4 \, \nabla^2 \Phi - 4 \big( \partial \Phi \big)^2 = \frac{3}{k} \, \frac{1 + \l^2}{1 - \l^2} \, ,
\\[5pt]
& \b_{g_{ab}} = R_{ab} + 2 \, \nabla_a \nabla_b \Phi = - \frac{2}{k} \frac{\l}{1 - \l^2} \, \eta_{ab} \, .
 \end{aligned}
\end{equation}
Here $R$ and $R_{ab}$ are the Ricci scalar and tensor respectively. The latter is computed in the frame \eqref{CS3lambdaFrame}, while $\eta_{ab} = \diag (-1 , +1 , +1 )$. As it is expected, $\b_{g_{ab}}$ vanishes when $\l = 0$. The frame components \eqref{CS3lambdaFrame} also satisfy the following relations
\begin{equation}
 \label{CS3lambdaFrameIdentities}
 d \big( e^{- \Phi} \, \fe^1 \big) = 0 \, , \qquad d \big( e^{- \Phi} \, \fe^1 \wedge \fe^3 \big) = d \big( e^{- \Phi} \, \fe^2 \wedge \fe^3 \big) = 0 \, .
\end{equation}
These will be particularly useful for constructing the RR fields.

The background fields for the non-compact deformed coset $\nicefrac{SO(1,3)_{-k}}{SO(3)_{-k}}$ can be obtained from those above by applying the analytic continuation
\begin{equation}
 \label{CompactToNonCompactAnalyticContinuation}
 \om \to i \om \, , \qquad k \to - k \, .
\end{equation}
The frame in this case reads
\begin{equation}
 \label{CH3lambdaFrame}
 \begin{aligned}
  & \hfe^1 = \frac{2 \, \l_-}{\sqrt{\cA}} \Big( \om \, x \, dx + \frac{y}{\om} \, dy - \frac{\cA}{\om^2_-} \, d\om \Big) \, ,
  \\[5pt]
  & \hfe^2 = \frac{2 \, \l_+}{\sqrt{\cA \, \cD}} \Big( \cD \, \om \, dx - \frac{x \, y}{\om} \, dy + \frac{\cA \, x}{\om^2_-} \, d\om \Big) \, ,
  \\[5pt]
  & \hfe^3 = \frac{2 \, \l_+}{\sqrt{\cD}} \Big( \frac{dy}{\om} + \frac{y}{\om^2_-} \, d\om \Big) \, .
 \end{aligned}
\end{equation}
The space has again positive signature $(+ , + , +)$. The analytic continuation implies that the scalar $\Phi$ is given by
\begin{equation}
 e^{- 2 \Phi} = - 8 \, \frac{\cA \, \om^2}{\om^4_-} \, .
\end{equation}
The $\b$-functions $\b^{\Phi}$ and $\b_{g_{ab}}$ differ from \eqref{CS3lambdaBetaFunctions} by an overall sign. Also, the frame \eqref{CH3lambdaFrame} satisfies the relations \eqref{CS3lambdaFrameIdentities}. In the following sections, we will refer to this model as $CH_{3 , \l}$.

It is also worth mentioning that there is a consistent way to take the limit $\l \to 1$, in both models discussed above. As is explained in \cite{Sfetsos:2013wia,Demulder:2015lva}, this requires an appropriate zoom-in on the $SO(4)$ and $SO(1,3)$ group elements near the identity. The outcome of such a limit is the non-Abelian T-dual of the geometric cosets $\nicefrac{SO(4)}{SO(3)}$ and $\nicefrac{SO(1,3)}{SO(3)}$, respectively.

%%%%%%%%%%%%%%%%%%%%%%%%%%%%%%%%%%%%%%%%%%%%%%%%%%%%%%%%%%%%%%
\subsection{The $\nicefrac{SO(5)_k}{SO(4)_k}$ $\l$-deformed model}
\label{CS4Lambda}

Next, we consider a $\s$-model with a four-dimensional target space, specifically the $\l$-model on $\nicefrac{SO(5)_k}{SO(4)_k}$. The metric has positive signature $(+ , + , + , +)$ and can be written in terms of the frame
\begin{equation}
 \label{CS4lambdaFrame}
 \begin{aligned}
  & \fe^1 = - \frac{2 \, \l_-}{\sqrt{\cA \cB \cD}} \Bigg( \frac{\cB \cD}{\om y} dx + x \Big( \frac{\cD z^2}{\om y^2} + \om y^2 \Big) dy - \frac{\cA x z}{\om y} dz - \frac{\cA \cB x}{\om^2_+ y} d\om \Bigg) \, ,
  \\[5pt]
  & \fe^2 = \frac{2 \, \l_+}{\sqrt{\cA \cB}} \Bigg( \frac{\cB x}{\om y} dx + \Big( \frac{x^2 z^2}{\om y^2} - \om y^2 \Big) dy + \frac{\cA z}{\om y} dz + \frac{\cA \cB}{\om^2_+ y} d\om \Bigg) \, ,
  \\[5pt]
  & \fe^3 = \frac{2 \, \l_-}{\sqrt{\cB \cD}} \Big( \om \, y \, dy + \frac{z}{\om} dz + \frac{\cB}{\om^2_+} d\om \Big) \, ,
  \\[5pt]
  & \fe^4 = - 2 \, \l_+ \Big( \frac{dz}{\om \, y} - \frac{z}{\om^2_+ y} d\om \Big) \, .
 \end{aligned}
\end{equation}
The constants $\l_\pm$ are given in \eqref{LambdaPlusMinus} and the functions $(\om_\pm , \cA , \cD)$ in \eqref{OmegaPlusMinusAD}. We further define the function $\cB$ as
\begin{equation}
 \cB = y^2 - z^2 \, .
\end{equation}
Besides the metric, this model also comes with a scalar $\Phi$ which now reads
\begin{equation}
 \label{CS4lambdaScalar}
 e^{- 2 \Phi} = 64 \, \frac{\cA \, \cB \, \om^4}{\om^6_+} \, .
\end{equation}
In analogy to the previous example, the content of this model will be denoted as $CS^4_{\l}$.

The $\b$-functions for the scalar $\Phi$ and the metric now become
\begin{equation}
 \label{CS4lambdaBetaFunctions}
 \begin{aligned}
  & \b^{\Phi} = R + 4 \, \nabla^2 \Phi - 4 \big( \partial \Phi \big)^2 = \frac{6}{k} \, \frac{1 + \l^2}{1 - \l^2} \, ,
\\[5pt]
& \b_{g_{ab}} = R_{ab} + 2 \, \nabla_a \nabla_b \Phi  = \frac{3}{k} \frac{\l}{1 - \l^2} \, \bar{\eta}_{ab} \, ,
 \end{aligned}
\end{equation}
where now $\bar{\eta}_{ab} = \diag(+1 , -1 , +1 , -1)$. Clearly $\b_{g_{ab}}$ vanishes when $\l = 0$. Moreover, the frame components \eqref{CS4lambdaFrame} satisfy the relations below
\begin{equation}
 \label{CS4lambdaFrameIdentities}
 d \big( e^{- \Phi} \, \fe^1 \wedge \fe^3 \big) = d \big( e^{- \Phi} \, \fe^2 \wedge \fe^4 \big) = 0 \, , \qquad d \big( e^{- \Phi} \, \fe^1 \wedge \fe^3 \wedge \fe^4 \big) = 0 \, .
\end{equation}
Later on, we will take advantage of these relations in order to construct ansatze for the RR fields.

Like in the previous example, the non-compact version of $CS^4_{\l}$ can be obtained by applying the analytic continuation \eqref{CompactToNonCompactAnalyticContinuation}. This results in the $\l$-model on $\nicefrac{SO(1,4)_{-k}}{SO(4)_{-k}}$ which will be denoted as $CH_{4 , \l}$. The corresponding metric has positive signature $(+ , + , + , +)$ and it can be expressed in terms of the frame
\begin{equation}
 \label{CH4lambdaFrame}
 \begin{aligned}
  & \hfe^1 = - \frac{2 \, \l_-}{\sqrt{\cA \cB \cD}} \Bigg( \frac{\cB \cD}{\om y} dx + x \Big( \frac{\cD z^2}{\om y^2} - \om y^2 \Big) dy - \frac{\cA x z}{\om y} dz + \frac{\cA \cB x}{\om^2_- y} d\om \Bigg) \, ,
  \\[5pt]
  & \hfe^2 = \frac{2 \, \l_+}{\sqrt{\cA \cB}} \Bigg( \frac{\cB x}{\om y} dx + \Big( \frac{x^2 z^2}{\om y^2} + \om y^2 \Big) dy + \frac{\cA z}{\om y} dz - \frac{\cA \cB}{\om^2_- y} d\om \Bigg) \, ,
  \\[5pt]
  & \hfe^3 = - \frac{2 \, \l_-}{\sqrt{\cB \cD}} \Big( \om \, y \, dy - \frac{z}{\om} dz + \frac{\cB}{\om^2_-} d\om \Big) \, ,
  \\[5pt]
  & \hfe^4 = - 2 \, \l_+ \Big( \frac{dz}{\om \, y} + \frac{z}{\om^2_- y} d\om \Big) \, .
 \end{aligned}
\end{equation}
Also, for the scalar $\Phi$ we get
\begin{equation}
 \label{CH4lambdaScalar}
 e^{- 2 \Phi} = 64 \, \frac{\cA \, \cB \, \om^4}{\om^6_-} \, .
\end{equation}
The $\b$-functions $\b^{\Phi}$, $\b_{g_{ab}}$ differ from \eqref{CS4lambdaBetaFunctions} by an overall minus sign and the components \eqref{CH4lambdaFrame} still satisfy \eqref{CS4lambdaFrameIdentities}.

Finally, the non-Abelian T-duals of the geometric cosets $\nicefrac{SO(5)}{SO(4)}$ and $\nicefrac{SO(1,4)}{SO(4)}$ can be obtained in the limit $\l \to 1$. Consistency of the limit involves a zoom-in on the group elements of $SO(5)$ and $SO(1,4)$ near the identity, respectively.

%%%%%%%%%%%%%%%%%%%%%%%%%%%%%%%%%%%%%%%%%%%%%%%%%%%%%%%%%%%%%%
\section{Solutions with $CS^3_{\l}$}
\label{Section3}

We now turn to the construction of type-II supergravity solutions based on the $CS^3_{\l}$ model discussed in Section \ref{CS3Lambda}. In the next section, we will consider that the ten-dimensional spacetime assumes the direct product form $\cM_7 \times CS^3_{\l}$. Here $\cM_7$ is a seven-dimensional manifold with a metric expressed in terms of the frame $(e^0 , \ldots , e^6)$ as
\begin{equation}
 ds^2_{\cM_7} = - \big( e^0 \big)^2 + \big( e^1 \big)^2 + \ldots + \big( e^6 \big)^2 \, .
\end{equation}
The other three directions are identified as $e^7 \to \fe^1$, $e^8 \to \fe^2$ and $e^9 \to \fe^3$ with $(\fe^1 , \fe^2 , \fe^3)$ given in \eqref{CS3lambdaFrame}. Moreover, we will take the NS two-form to be zero and the dilaton to be given by \eqref{CS3lambdaScalar}. The curvature on $\cM_7$ can be already deduced by the fact that the ten-dimensional geometry has a direct product form and the first of the properties \eqref{CS3lambdaBetaFunctions}. Indeed, from the dilaton equation \eqref{DilatonEOM} we find
\begin{equation}
 \label{RicciScalarM7}
 R_{\cM_7} + \frac{3}{k} \, \frac{1 + \l^2}{1 - \l^2} = 0 \, .
\end{equation}
Here, $R_{\cM_7}$ is the Ricci scalar on $\cM_7$. Therefore, $\cM_7$ is a space of constant negative curvature. We can further constrain $\cM_7$ by introducing specific ansatze for the RR fields.

%%%%%%%%%%%%%%%%%%%%%%%%%%%%%%%%%%%%%%%%%%%%%%%%%%%%%%%%%%%%%%
\subsection{Type-IIA backgrounds on $\cM^t_4 \times \cM_3 \times CS^3_{\l}$}
\label{Mt4M3CS3lambdaIIA}

Starting with solutions of the type-IIA supergravity, we adopt the following ansatz for the RR sector
\begin{equation}
 \label{AnsatzM4tM3IIA}
 F_0 = 0 \, , \qquad F_2 = 2 \, c_1 \, e^{- \Phi} \fe^2 \wedge \fe^3 \, , \qquad F_4 = 2 \, c_2 \, e^{- \Phi} e^4 \wedge e^5 \wedge e^6 \wedge \fe^1 \, ,
\end{equation}
where $c_1$ and $c_2$ are constants. From \eqref{CS3lambdaFrameIdentities} one can easily verify that the Bianchi equation \eqref{BianchisIIA} for $F_2$ is satisfied. Moreover, the Bianchi and flux equations \eqref{BianchisIIA} and \eqref{FluxesIIA} for $F_4$ imply
\begin{equation}
 d \big( e^0 \wedge e^1 \wedge e^2 \wedge e^3 \big) = 0 \, , \qquad d \big( e^4 \wedge e^5 \wedge e^6 \big) = 0 \, .
\end{equation}
The above guarantee that also the flux equation \eqref{FluxesIIA} for $F_2$ holds true. On top of that, it indicates that $\cM_7$ splits into the direct product $\cM_7 = \cM^t_4 \times \cM_3$. In particular, $\cM^t_4$ is a four-dimensional space spanned by $(e^0 , e^1 , e^2 , e^3)$ and $\cM_3$ a three-dimensional space spanned by $(e^4 , e^5 , e^6)$. The superscript in $\cM^t_4$ means that the time direction sits in the four-dimensional space.

Having the Bianchi and flux equations under control, we continue with the Einstein equations \eqref{EinsteinIIA}. The non-zero components of $\cT^{IIA}_{ab}$ in the frame $e^a \, (a = 0 , 1 , \ldots , 9)$ are
\begin{equation}
 \label{TddM7IIA}
 \begin{aligned}
  & \cT^{IIA}_{ab} = - \big( c^2_1 + c^2_2 \big) \hat{\eta}_{ab} \, , \qquad a, b = 0 , 1 , 2 , 3 \, ,
  \\[5pt]
  & \cT^{IIA}_{ab} = \big( c^2_2 - c^2_1 \big) \d_{ab} \, , \qquad\quad a, b = 4 , 5 , 6 \, ,
  \\[5pt]
  & \cT^{IIA}_{ab} = \big( c^2_1 - c^2_2 \big) \eta_{ab} \, , \qquad\quad a, b = 7 , 8 , 9 \, ,
 \end{aligned}
\end{equation}
where $\hat{\eta}_{ab} = \diag (-1 , +1 , +1 , +1)$ and $\eta_{ab} = \diag (-1 , +1 , +1)$. The first two lines above together with \eqref{EinsteinIIA} imply that the Ricci tensor on $\cM^t_4$ and $\cM_3$ can be expressed in terms of the constants $c_1$ and $c_2$ as
\begin{equation}
 \label{RddM7IIA}
 \begin{aligned}
  & \cM^t_4: \qquad R_{ab} = - \big( c^2_1 + c^2_2 \big) \hat{\eta}_{ab} \, , \qquad a, b = 0 , 1 , 2 , 3 \, ,
  \\[5pt]
  & \cM_3: \qquad R_{ab} = \big( c^2_2 - c^2_1 \big) \d_{ab} \, , \qquad\quad a, b = 4 , 5 , 6 \, .
 \end{aligned}
\end{equation}
From this we can compute the Ricci scalar on $\cM_7$ and combine it with \eqref{RicciScalarM7}. This results into the following condition for the constants $c_1$ and $c_2$
\begin{equation}
 \label{paramsM4tM31}
 7 \, c^2_1 + c^2_2 = \frac{3}{k} \, \frac{1 + \l^2}{1 - \l^2} \, .
\end{equation}
A second condition for $c_1$ and $c_2$ can be found by combining the third line of \eqref{TddM7IIA} with \eqref{EinsteinIIA} and \eqref{CS3lambdaBetaFunctions}
\begin{equation}
 \label{paramsM4tM32}
 c^2_2 - c^2_1 = \frac{2}{k} \frac{\l}{1 - \l^2} \, .
\end{equation}
The equations \eqref{paramsM4tM31} and \eqref{paramsM4tM32} can be solved for
\begin{equation}
 \label{solutionM4tM3IIA}
 c^2_1 = \frac{1}{8 \, k} \frac{3 - 2 \, \l + 3 \l^2}{1 - \l^2} \, , \qquad c^2_2 = \frac{1}{8 \, k} \frac{3 + 14 \, \l + 3 \l^2}{1 - \l^2} \, .
\end{equation}
Obviously $c^2_1 , \, c^2_2 \geq 0$ for positive $k$ and arbitrary values of $\l$ in the interval $[0 , 1)$. Hence the supergravity backgrounds are real without imposing any restriction on $\l$. The above values of the parameters $c_1$ and $c_2$ also imply that the Ricci tensor components on $\cM^t_4$ and $\cM_3$ are respectively
\begin{equation}
 \label{RicciMt4M3}
 \begin{aligned}
  & \cM^t_4: \qquad R_{ab} = - \frac{3}{4 k} \frac{1 + \l}{1 - \l} \hat{\eta}_{ab} \, , \qquad a, b = 0 , 1 , 2 , 3 \, ,
  \\[5pt]
  & \cM_3: \qquad R_{ab} = \frac{2}{k} \frac{\l}{1 - \l^2} \d_{ab} \, , \qquad\quad a, b = 4 , 5 , 6 \, .
 \end{aligned}
\end{equation}
This indicates that $\cM_7$ decomposes into the direct product of two Einstein spaces. The space $\cM^t_4$ has constant negative curvature, while $\cM_3$ has constant positive curvature. Notice that when $\l = 0$ the space $\cM_3$ is flat.

\vskip 10pt

\noindent\textbf{Comments}

\vskip 10pt

\begin{itemize}
 \item In summary, the geometry of the ten-dimensional background is described by a metric of the form
 \begin{equation}
  \label{metricMt4M3IIA}
  ds^2 = ds^2_{\cM^t_4} + ds^2_{\cM_3} + \big( \fe^1 \big)^2 + \big( \fe^2 \big)^2 + \big( \fe^3 \big)^2 \, ,
 \end{equation}
 where  $\fe^a \, (a = 1 , 2 , 3)$ is the frame given in \eqref{CS3lambdaFrame}, $ds^2_{\cM^t_4}$ is the line-element of the four-dimensional space $\cM^t_4$ and $ds^2_{\cM_3}$ that of $\cM_3$. The spaces $\cM^t_4$ and $\cM_3$ are spanned by the frames $(e^0 , e^1 , e^2 , e^3)$ and $(e^4 , e^5 , e^6)$ respectively. Also, they are normalised according to \eqref{RicciMt4M3}.
 
 A natural choice for $\cM^t_4$ is $AdS_4$. A second option for $\cM^t_4$ is the direct product of $AdS_2$ and a two-dimensional hyperbolic plane $H_2$, i.e. $AdS_2 \times H_2$. The three-dimensional space $\cM_3$ can be identified with a three-sphere, $S^3$.
 
  \item The rest of the supergravity fields consist of the dilaton $\Phi$ given in \eqref{CS3lambdaScalar}, and the RR sector described in \eqref{AnsatzM4tM3IIA}. The constants $c_1$, $c_2$ take the values indicated in \eqref{solutionM4tM3IIA} with $\l \in [0 , 1)$.

 \item It is also possible to attempt constructing backgrounds based on the target space of the $CH_{3, \l}$ model, instead of $CS^3_{\l}$, using an ansatz similar to \eqref{AnsatzM4tM3IIA}. In this case, the $\b$-functions $\b^{\Phi}$ and $\b_{g_{ab}}$ differ from \eqref{CS3lambdaBetaFunctions} by an overall sign. As a result, equations \eqref{paramsM4tM31} and \eqref{paramsM4tM32} would come with the opposite sign in the right hand side. This means that \eqref{paramsM4tM31} would not allow for a real solution in terms of $c_1$ and $c_2$. Therefore, backgrounds on $\cM_7 \times CH_{3, \l}$ are not of our interest. 
\end{itemize}

%%%%%%%%%%%%%%%%%%%%%%%%%%%%%%%%%%%%%%%%%%%%%%%%%%%%%%%%%%%%%%
\subsection{Type-IIB backgrounds on $\cM^t_3 \times \cM_4 \times CS^3_{\l}$}
\label{Mt3M4CS3lambdaIIA}

We now propose the following ansatz for the RR forms
\begin{equation}
 \label{AnsatzM3tM4IIB}
 F_1 = 2 \, c_1 \, e^{- \Phi} \, \fe^1 \, , \qquad F_3 = 0 \, , \qquad F_5 = 2 \, c_2 \, e^{- \Phi} \big( 1 + \star \big) e^3 \wedge e^4 \wedge e^5 \wedge e^6 \wedge \fe^1 \, ,
\end{equation}
where $c_1$ and $c_2$ are constants. Using \eqref{CS3lambdaFrameIdentities} it is easy to convince ourselves that the Bianchi identities \eqref{BianchisIIB} for $F_1$ and $F_3$ are satisfied, while that for $F_5$ implies
\begin{equation}
 d \big( e^0 \wedge e^1 \wedge e^2 \big) = d \big( e^3 \wedge e^4 \wedge e^5 \wedge e^6 \big) = 0 \, .
\end{equation}
The last guarantee that also \eqref{FluxesIIB} are satisfied. The above conditions also allow us to interpret $e^0 \wedge e^1 \wedge e^2$ and $e^3 \wedge \ldots \wedge e^6$ as the volume forms on a three- and a four-dimensional space respectively. This suggests that $\cM_7$ can be written as the direct product of the aforementioned subspaces, i.e. $\cM_7 = \cM^t_3 \times \cM_4$.

Moving to the Einstein equations \eqref{EinsteinIIB}, we compute the non-vanishing entries of $\cT^{IIB}_{ab}$ in the frame $e^a \, (a = 0 , 1 , \ldots , 9)$
\begin{equation}
 \label{TddM7IIB}
 \begin{aligned}
  & \cT^{IIB}_{ab} = - \big( c^2_1 + c^2_2 \big) \eta_{ab} \, , \qquad a, b = 0 , 1 , 2 \, ,
  \\[5pt]
  & \cT^{IIB}_{ab} = \big( c^2_2 - c^2_1 \big) \d_{ab} \, , \qquad\quad a, b = 3 , 4 , 5 , 6 \, ,
  \\[5pt]
  & \cT^{IIB}_{ab} = - \big( c^2_1 + c^2_2 \big) \eta_{ab} \, , \qquad a, b = 7 , 8 , 9 \, ,
 \end{aligned}
\end{equation}
where $\eta_{ab} = \diag (-1 , +1 , +1)$. From the first two lines above and the equations of motion \eqref{EinsteinIIB} we can read the Ricci tensor on $\cM_7$
\begin{equation}
 \label{RddM7IIB}
 \begin{aligned}
  & \cM^t_3: \qquad R_{ab} = - \big( c^2_1 + c^2_2 \big) \eta_{ab} \, , \qquad a, b = 0 , 1 , 2 \, ,
  \\[5pt]
  & \cM_4: \qquad R_{ab} = \big( c^2_2 - c^2_1 \big) \d_{ab} \, , \qquad\quad a, b = 3 , 4 , 5 , 6 \, .
 \end{aligned}
\end{equation}
If we combine this with \eqref{RicciScalarM7} we obtain the following relation for the constants $c_1$ and $c_2$
\begin{equation}
 \label{paramsM3tM41}
 7 c^2_1 - c^2_2 = \frac{3}{k} \, \frac{1 + \l^2}{1 - \l^2} \, .
\end{equation}
A second relation can be obtained from the third line of \eqref{TddM7IIB} and the second property in \eqref{CS3lambdaBetaFunctions} in view of \eqref{EinsteinIIB}. Indeed this gives
\begin{equation}
 \label{paramsM3tM42}
 c^2_1 + c^2_2 = \frac{2}{k} \frac{\l}{1 - \l^2} \, .
\end{equation}
Solving the last two for $c_1$ and $c_2$ we find
\begin{equation}
 \label{solutionM3tM4IIB}
 c^2_1 = \frac{1}{8 \, k} \frac{3 + 2 \, \l + 3 \l^2}{1 - \l^2} \, , \qquad c^2_2 = - \frac{1}{8 \, k} \frac{3 - 14 \, \l + 3 \l^2}{1 - \l^2} \, .
\end{equation}
Requiring that the solution is real, i.e. $c^2_1 , c^2_2 \geq 0$, restricts the allowed values of $\l$ to
\begin{equation}
 \label{lambdaBoundM3tM4IIB}
 \frac{7 - 2 \sqrt{10}}{3} \leq \l < 1 \, .
\end{equation}
Notice that for these values of the parameters $c_1$ and $c_2$, the submanifolds spanned by the directions $(e^0,e^1,e^2)$ and $(e^3,e^4,e^5,e^6)$ are Einstein spaces of constant negative curvature. In particular, they are scaled such that
\begin{equation}
 \label{RicciMt3M4}
 \begin{aligned}
  & \cM^t_3: \qquad R_{ab} = - \frac{2}{k} \frac{\l}{1 - \l^2} \eta_{ab} \, , \qquad\quad a, b = 0 , 1 , 2 \, ,
  \\[5pt]
  & \cM_4: \qquad R_{ab} = - \frac{3}{4 k} \frac{1 - \l}{1 + \l} \d_{ab} \, , \qquad\quad a, b = 3 , 4 , 5 , 6 \, .
 \end{aligned}
\end{equation}

\vskip 10pt

\noindent\textbf{Comments}

\vskip 10pt

\begin{itemize}
 \item The geometry of the supergravity solution discussed above is given by the line element
 \begin{equation}
  ds^2 = ds^2_{\cM^t_3} + ds^2_{\cM_4} + \big( \fe^1 \big)^2 + \big( \fe^2 \big)^2 + \big( \fe^3 \big)^2 \, ,
 \end{equation}
 with $\fe^a \, (a = 1 , 2 , 3)$ being the frame given in  \eqref{CS3lambdaFrame}. Here, $ds^2_{\cM^t_3}$ is the line element of a three-dimensional space $\cM^t_3$ spanned by the frame $(e^0 , e^1 , e^2)$, and $ds^2_{\cM_4}$ the line element of $\cM_4$ spanned by $(e^3 , e^4 , e^5 , e^6)$. The spaces $\cM^t_3$ and $\cM_4$ are normalised according to \eqref{RicciMt3M4}.
 
 One can take $\cM^t_3$ to be an $AdS_3$, while for $\cM_4$ one can consider for example a four-dimensional hyperbolic space $H_4$ or the direct product of two hyperbolic planes $H_2 \times \tilde{H}_2$.
 
 \item The ten-dimensional geometry above is also supported by a dilaton $\Phi$ given in \eqref{CS3lambdaScalar} and the set of RR fields \eqref{AnsatzM3tM4IIB}. The constants $c_1$ and $c_2$ now take the values \eqref{solutionM3tM4IIB} and $\l$ is bounded as in \eqref{lambdaBoundM3tM4IIB}.
 
 \item Notice that the ansatz \eqref{AnsatzM3tM4IIB} would not work for $CH_{3, \l}$ in place of $CS^3_{\l}$. The reason is that in this case, the right hand side of \eqref{paramsM3tM41} and \eqref{paramsM3tM42} would change by an overall minus sign. Therefore, equation \eqref{paramsM3tM42} would imply that at least one of the parameters $c_1$ or $c_2$ is imaginary. 
\end{itemize}

%%%%%%%%%%%%%%%%%%%%%%%%%%%%%%%%%%%%%%%%%%%%%%%%%%%%%%%%%%%%%%
\subsection{Type-IIB backgrounds on $\cM^t_4 \times \cM_3 \times CS^3_{\l}$}
\label{Mt4M3CS3lambdaIIB}

We can also derive type-IIB backgrounds by applying a double analytic continuation in the construction discussed above. The effect of this action is that the seven-dimensional manifold $\cM_7$ now splits as $\cM^t_4 \times \cM_3$, and the time direction now sits in the four-dimensional subspace. Equivalently, this amounts to considering the ansatz
\begin{equation}
 \label{AnsatzM4tM3IIB}
 F_1 = 2 \, c_1 \, e^{- \Phi} \, \fe^1 \, , \qquad F_3 = 0 \, , \qquad F_5 = 2 \, c_2 \, e^{- \Phi} \big( 1 + \star \big) e^0 \wedge e^1 \wedge e^2 \wedge e^3 \wedge \fe^1 \, ,
\end{equation}
where again $c_1$ and $c_2$ are constants to be determined. In this case the Bianchi and flux equations \eqref{BianchisIIB} and \eqref{FluxesIIB} imply
\begin{equation}
 d \big( e^0 \wedge e^1 \wedge e^2 \wedge e^3 \big) = d \big( e^4 \wedge e^5 \wedge e^6 \big) = 0 \, .
\end{equation}
The rest of the computation goes along the same lines as in the previous construction. Nevertheless, the outcome differs only by an overall sign in $c^2_2$. In particular
\begin{equation}
 \label{solutionM4tM3IIB}
 c^2_1 = \frac{1}{8 \, k} \frac{3 + 2 \, \l + 3 \l^2}{1 - \l^2} \, , \qquad c^2_2 = \frac{1}{8 \, k} \frac{3 - 14 \, \l + 3 \l^2}{1 - \l^2} \, .
\end{equation}
The solution is real when $\l$ takes values in
\begin{equation}
 \label{lambdaBoundM4tM3IIB}
 0 \leq \l \leq \frac{7 - 2 \sqrt{10}}{3} \, .
\end{equation}
The Ricci tensor on the subspaces $\cM^t_4$ and $\cM_3$ now reads
\begin{equation}
 \label{RicciMt4M3IIB}
 \begin{aligned}
  & \cM^t_4: \qquad R_{ab} = - \big( c^2_1 + c^2_2 \big) \hat{\eta}_{ab} = -  \frac{3}{4 k} \frac{1 - \l}{1 + \l} \hat{\eta}_{ab} \, , \qquad a, b = 0 , 1 , 2 , 3 \, ,
  \\[5pt]
  & \cM_3: \qquad R_{ab} = - \big( c^2_1 - c^2_2 \big) \d_{ab} = - \frac{2}{k} \frac{\l}{1 - \l^2} \d_{ab} \, , \qquad a, b = 4 , 5 , 6 \, .
 \end{aligned}
\end{equation}
%
%In this case $\eta_{ab} = \diag (-1 , +1 , +1 , +1)$. 
Clearly, both $\cM^t_4$ and $\cM_3$ are Einstein spaces of negative curvature. Notice that when $\l = 0$ the space $\cM_3$ is flat and therefore, we can safely take it to be a three-dimensional torus $T^3$.

\vskip 10pt

\noindent\textbf{Comments}

\vskip 10pt

\begin{itemize}
 \item The metric of this background has the same form as \eqref{metricMt4M3IIA}. Therefore, one can again choose $\cM^t_4$ to be $AdS_4$ or $AdS_2 \times H_2$. However, now $\cM_3$ has negative curvature and as such it can be identified with a three-dimensional hyperbolic space, $H_3$. The spaces $\cM^t_4$ and $\cM_3$ are normalised according to \eqref{RicciMt4M3IIB}.
 
 \item The supergravity solution also contains a dilaton $\Phi$ given in \eqref{CS3lambdaScalar}. The RR sector consists of the forms $F_1$ and $F_5$ written in \eqref{AnsatzM4tM3IIB}. Now, the constants $c_1$ and $c_2$ take values according to \eqref{solutionM4tM3IIB}, with $\l$ restricted as in \eqref{lambdaBoundM4tM3IIB}.
 
 \item Similarly to the case of the previous section, if we replace $CS^3_{\l}$ by $CH_{3, \l}$, we can not have a real solution corresponding to the ansatz \eqref{AnsatzM4tM3IIB}. 
\end{itemize}

%%%%%%%%%%%%%%%%%%%%%%%%%%%%%%%%%%%%%%%%%%%%%%%%%%%%%%%%%%%%%%
\section{Solutions with $CS^4_{\l}$ and $CH_{4, \l}$}
\label{Section4}

In this section we construct type-II backgrounds by uplifting the $CS^4_{\l}$ model in ten dimensions. Therefore, the four-dimensional target space of $CS^4_{\l}$, described in Sec. \ref{CS4Lambda}, must be completed with a six-dimensional manifold. Hence, for the ten-dimensional spacetime we will assume the decomposition into the direct product $\cM_6 \times CS^4_{\l}$. We will express the metric on $\cM_6$ in terms of the frame $(e^0 , \ldots , e^5)$ as
\begin{equation}
 \label{MetricM6}
 ds^2_{\cM_6} = - \big( e^0 \big)^2 + \big( e^1 \big)^2 + \ldots + \big( e^5 \big)^2 \, .
\end{equation}
The other directions are taken to be those in \eqref{CS4lambdaFrame}. In particular, we identify $e^6 \to \fe^1$, $e^7 \to \fe^2$, $e^8 \to \fe^3$ and $e^9 \to \fe^4$. Moreover, for the other NS fields we take the two-form to be trivial and the dilaton to be given by \eqref{CS4lambdaScalar}. We can extract the curvature on $\cM_6$ using the dilaton equation \eqref{DilatonEOM} in conjunction with the first from \eqref{CS4lambdaBetaFunctions}. Indeed, for the Ricci scalar on $\cM_6$ we find
\begin{equation}
 \label{RicciScalarM6}
 R_{\cM_6} + \frac{6}{k} \, \frac{1 + \l^2}{1 - \l^2} = 0 \, .
\end{equation}
Therefore, $\cM_6$ is a space of constant negative curvature. In what follows we will see that choosing an ansatz for the RR sector restricts further the geometry of $\cM_6$.

%%%%%%%%%%%%%%%%%%%%%%%%%%%%%%%%%%%%%%%%%%%%%%%%%%%%%%%%%%%%%%
\subsection{Type-IIA backgrounds on $\cM_6 \times CS^4_{\l}$}
\label{M6CS4lambdaIIA}

Focusing on solutions of the type-IIA supergravity, the first ansatz that we would like to propose for the RR sector reads
\begin{equation}
 \label{AnsatzM6tIIA}
 F_0 = 0 \, , \qquad F_2 = 2 \, e^{- \Phi} \big( c_1 \, \fe^1 \wedge \fe^3 + c_2 \, \fe^2 \wedge \fe^4 \big) \, , \qquad F_4 = 0 \, .
\end{equation}
It is clear that with this choice, the Bianchi identities \eqref{BianchisIIA} are automatically satisfied in view of \eqref{CS4lambdaFrameIdentities}. Moving to the flux equations \eqref{FluxesIIA}, we only need to take into account that for $F_2$, which implies
\begin{equation}
 d \big( e^0 \wedge e^1 \wedge e^2 \wedge e^3 \wedge e^4 \wedge e^5 \big) = 0.
\end{equation}
The last tells us that $e^0 \wedge e^1 \wedge e^2 \wedge e^3 \wedge e^4 \wedge e^5$ can be seen as the volume form on $\cM_6$.

We can infer more about the geometry of $\cM_6$ by looking at the Einstein equations \eqref{EinsteinIIA}. Computing the non-vanishing components of $\cT^{IIA}_{ab}$ in the frame $e^a \, (a = 0 , 1 , \ldots , 9)$ we find
\begin{equation}
 \label{TddAdS6IIA}
 \begin{aligned}
  & \cT^{IIA}_{ab} = - \big( c^2_1 + c^2_2 \big) \dot{\eta}_{ab} \, , \qquad a, b = 0 , 1 , 2 , 3 , 4 , 5 \, ,
  \\[5pt]
  & \cT^{IIA}_{ab} = \big( c^2_1 - c^2_2 \big) \d_{ab} \, , \qquad\quad a, b = 6 , 7 , 8 , 9 \, ,
 \end{aligned}
\end{equation}
where $\dot{\eta}_{ab} = \diag (-1 , +1 , +1 , +1 , +1 , +1)$. The first line above, together with the Einstein equations \eqref{EinsteinIIA} indicate that the Ricci tensor on $\cM_6$ is
\begin{equation}
 R_{ab} = - \big( c^2_1 + c^2_2 \big) \dot{\eta}_{ab} \, , \qquad a, b = 0 , 1 , 2 , 3 , 4 , 5 \, .
\end{equation}
We can now compute the Ricci scalar on $\cM_6$ in terms of the constants $c_1$, $c_2$ and combine it with \eqref{RicciScalarM6}. Doing so, we find
\begin{equation}
 \label{paramsM6tIIA1}
 c^2_1 + c^2_2 = \frac{1}{k} \, \frac{1 + \l^2}{1 - \l^2} \, .
\end{equation}
Another constraint for $c_1$ and $c_2$ can be derived from the second line in \eqref{TddAdS6IIA} when combined with the second line in \eqref{CS4lambdaBetaFunctions} and \eqref{EinsteinIIA}. This gives
\begin{equation}
 \label{paramsM6tIIA2}
 c^2_1 - c^2_2 = \frac{3}{k} \frac{\l}{1 - \l^2} \, .
\end{equation}
We can now easily solve \eqref{paramsM6tIIA1} and \eqref{paramsM6tIIA2} with respect to $c_1$ and $c_2$, where
\begin{equation}
 \label{solutionM6tIIA}
 c^2_1 = \frac{1}{2 \, k} \frac{1 + 3 \, \l + \l^2}{1 - \l^2} \, , \qquad c^2_2 = \frac{1}{2 \, k} \frac{1 - 3 \, \l + \l^2}{1 - \l^2}
\end{equation}
The supergravity background is real provided that $c^2_1 \, , c^2_2 \geq 0$. As a result, the deformation parameter $\l$ must be bounded as
\begin{equation}
 \label{lambdaBoundM6tIIA}
 0 \leq \l \leq \frac{3 - \sqrt{5}}{2} \, .
\end{equation}
Finally, the Ricci tensor on $\cM_6$ in terms of $\l$ is
\begin{equation}
 \label{RicciMt6}
 \cM_6: \qquad R_{ab} = - \frac{1}{k} \, \frac{1 + \l^2}{1 - \l^2} \, \dot{\eta}_{ab} \, , \qquad a, b = 0 , 1 , 2 , 3 , 4 , 5 \, .
\end{equation}
Therefore, $\cM_6$ is an Einstein space of negative constant curvature.

\vskip 10pt

\noindent\textbf{Comments}

\vskip 10pt

\begin{itemize}
 \item The ten-dimensional geometry in this case is the direct product of a six-dimensional manifold $\cM_6$ and the target space of the $CS^4_{\l}$ model. Therefore, its line element takes the form
 \begin{equation}
  ds^2 = ds^2_{\cM_6} + \big( \fe^1 \big)^2 + \big( \fe^2 \big)^2 + \big( \fe^3 \big)^2 + \big( \fe^4 \big)^2 \, ,
 \end{equation}
 where $\fe^a \, (a = 1 , \ldots , 4)$ is the frame given in \eqref{CS4lambdaFrame}. The space $\cM_6$ is spanned by the frame $(e^0 , \ldots , e^5)$ and is normalised according to \eqref{RicciMt6}.
 
 One can safely choose $\cM_6$ to be $AdS_6$. Nevertheless, this is not the only option that can be considered. Among the other possibilities are the direct products $AdS_2 \times H_2 \times \tilde{H}_2$, $AdS_2 \times H_4$, $AdS_3 \times H_3$ and $AdS_4 \times H_2$.
 
 \item Besides the metric, the solution contains the dilaton $\Phi$ whose expression is \eqref{CS4lambdaScalar}, and also a RR two-form given by \eqref{AnsatzM6tIIA}. The constants $c_1$ and $c_2$ are related to the deformation $\l$ through \eqref{solutionM6tIIA}, where $\l$ lies in the interval \eqref{lambdaBoundM6tIIA}.
 
 \item A similar construction with $CH_{4, \l}$ instead of $CS^4_{\l}$ would not be allowed. The reason is that the $\b$-functions $\b^\Phi$ and $\b_{g_{ab}}$ for $CH_{4 , \l}$ come with the opposite sign compared to the ones for $CS^4_{\l}$ in \eqref{CS4lambdaBetaFunctions}. That would amount to changing the sign in the right hand side of \eqref{paramsM6tIIA1} and \eqref{paramsM6tIIA2}. Hence, the solution for $c_1$ and $c_2$ would not be real. 
\end{itemize}

%%%%%%%%%%%%%%%%%%%%%%%%%%%%%%%%%%%%%%%%%%%%%%%%%%%%%%%%%%%%%%
\subsection{Type-IIA backgrounds on $\cM^t_2 \times \cM_4 \times CS^4_{\l}$}
\label{Mt2M4CS4lambdaIIA}

Continuing with solutions of the type-IIA supergravity, we consider the ansatz for the RR sector below
\begin{equation}
 \label{AnsatzM2tM4IIA}
 \begin{aligned}
  & F_0 = 0 \, ,
  \\[5pt]
  & F_2 = 2 \, e^{- \Phi} \big( c_1 \, \fe^1 \wedge \fe^3 + c_2 \, \fe^2 \wedge \fe^4 \big) \, ,
  \\[5pt]
  & F_4 = 2 \, e^{- \Phi} e^0 \wedge e^1 \wedge \big( c_3 \, \fe^1 \wedge \fe^3 + c_4 \, \fe^2 \wedge \fe^4 \big) \, ,
 \end{aligned}
\end{equation}
where $c_i \, (i = 1 , \ldots , 4)$ are constants. With this choice, one can verify that in view of \eqref{CS4lambdaFrameIdentities}, the Bianchi \eqref{BianchisIIA} and flux \eqref{FluxesIIA} equations for $F_4$ imply respectively
\begin{equation}
 \label{FormsCondsIIAM6}
 d \big( e^0 \wedge e^1 \big) = d \big( e^2 \wedge e^3 \wedge e^4 \wedge e^5 \big) = 0 \, .
\end{equation}
The rest of the Bianchi and flux equations are trivially satisfied except for the first of \eqref{FluxesIIA}. From this we get the condition
\begin{equation}
 \label{paramsM2tM4IIA1}
 c_1 \, c_3 + c_2 \, c_4 = 0 \, .
\end{equation}

More constraints for the constants $c_i \, (i = 1 , \ldots , 4)$ can be obtained by analysing the Einstein equations \eqref{EinsteinIIA}. Computing the non-zero entries of the matrix $ \cT^{IIA}_{ab}$ in the frame $e^a \, (a = 0 , 1 , \ldots , 9)$ we find
\begin{equation}
 \label{TddM6IIA}
 \begin{aligned}
  & \cT^{IIA}_{ab} = - \big( c^2_1 + c^2_2 + c^2_3 + c^2_4 \big) \tilde{\eta}_{ab} \, , \qquad\, a, b = 0 , 1 \, ,
  \\[5pt]
  & \cT^{IIA}_{ab} = \big( c^2_3 + c^2_4 - c^2_1 - c^2_2 \big) \d_{ab} \, , \qquad\quad a, b = 2 , 3 , 4 , 5 \, ,
  \\[5pt]
  & \cT^{IIA}_{ab} = \big( c^2_1 - c^2_2 - c^2_3 + c^2_4 \big) \bar{\eta}_{ab} \, , \qquad\quad a, b = 6 , 7 , 8 , 9 \, ,
 \end{aligned}
\end{equation}
where now $\tilde{\eta}_{ab} = \diag (-1 , +1)$. The first two lines above, together with the Einstein equations \eqref{EinsteinIIA}, suggest that $\cM_6$ factorises as $\cM^t_2 \times \cM_4$, where $\cM^t_2$ is spanned by $(e^0 , e^1)$ and $\cM_4$ by $(e^2 , e^3 , e^4 , e^5)$. In particular, for the Ricci tensors on $\cM^t_2$ and $\cM_4$ we find respectively
\begin{equation}
 \label{RddM6IIA}
 \begin{aligned}
  & \cM^t_2: \qquad R_{ab} = - \big( c^2_1 + c^2_2 + c^2_3 + c^2_4 \big) \tilde{\eta}_{ab} \, , \qquad a, b = 0 , 1 \, ,
  \\[5pt]
  & \cM_4: \qquad R_{ab} = \big( c^2_3 + c^2_4 - c^2_1 - c^2_2 \big) \d_{ab} \, , \qquad\quad a, b = 2 , 3 , 4 , 5 \, .
 \end{aligned}
\end{equation}
The factorisation $\cM_6 = \cM^t_2 \times \cM_4$ is also indicated by \eqref{FormsCondsIIAM6}. We can now compute the Ricci scalar on $\cM_6$ in terms of the constants $c_i \, (i = 1 , \ldots , 4)$. Doing so, and using \eqref{RicciScalarM6}, we find
\begin{equation}
 \label{paramsM2tM4IIA2}
 6 \, c^2_1 + 6 \, c^2_2 - 2 \, c^2_3 - 2 \, c^2_4 = \frac{6}{k} \, \frac{1 + \l^2}{1 - \l^2} \, .
\end{equation}
Also, the third line in \eqref{TddM6IIA} combined with the second line of \eqref{CS4lambdaBetaFunctions} and the Einstein equations \eqref{EinsteinIIA} implies
\begin{equation}
 \label{paramsM2tM4IIA3}
 c^2_1 - c^2_2 - c^2_3 + c^2_4 = \frac{3}{k} \frac{\l}{1 - \l^2} \, .
\end{equation}

In total, there are three independent algebraic constraints to solve with respect to the four constant parameters $c_i \, (i = 1 , \ldots , 4)$. More specifically, the equations we have to solve are \eqref{paramsM2tM4IIA1}, \eqref{paramsM2tM4IIA2}, and \eqref{paramsM2tM4IIA3}. Notice that all of them are invariant under
\begin{equation}
 \label{Symmetryc1c2c3c4}
 c_1 \leftrightarrow c_2 \, , \qquad c_3 \leftrightarrow c_4 \, , \qquad \l \to - \l \, .
\end{equation}
The above symmetry should also be respected by the solution of \eqref{paramsM2tM4IIA1}, \eqref{paramsM2tM4IIA2} and \eqref{paramsM2tM4IIA3}. We can solve the aforementioned conditions assuming either $c_1$ or $c_3$ to be arbitrary. Then the remaining solutions can be obtained applying the symmetry \eqref{Symmetryc1c2c3c4}. Here, we do not intent to solve \eqref{paramsM2tM4IIA1}, \eqref{paramsM2tM4IIA2} and \eqref{paramsM2tM4IIA3} in full generality. We will rather focus on the special cases explained below.

\subsubsection{Case $\mathbf{1}$: $\mathbf{c_4 = - c_1 , \, c_3 = c_2}$}
\label{Mt4T4IIA1}

Taking $c_4 = - c_1$ and $c_3 = c_2$ verifies \eqref{paramsM2tM4IIA1}. On the other hand, the equations \eqref{paramsM2tM4IIA2} and \eqref{paramsM2tM4IIA3} become
\begin{equation}
 c^2_1 + c^2_2 = \frac{3}{2 \, k} \, \frac{1 + \l^2}{1 - \l^2} \, , \qquad c^2_1 - c^2_2 = \frac{3}{2 \, k} \frac{\l}{1 - \l^2} \, .
\end{equation}
It is now easy to find the solution for $c_1$, $c_2$, $c_3$ and $c_4$, where
\begin{equation}
 \label{solutionM2tM4IIA1}
 c^2_1 = c^2_4 = \frac{3}{4 \, k} \frac{1 + \l + \l^2}{1 - \l^2} \, , \qquad  c^2_2 = c^2_3 = \frac{3}{4 \, k} \frac{1 - \l + \l^2}{1 - \l^2} \, .
\end{equation}
It turns out that the parameters $c_i \, (i = 1 , 2 , 3 , 4)$ are real for all the allowed values of the deformation $\l$. The Ricci tensor on $\cM^t_2$ and $\cM_4$ is respectively
\begin{equation}
 \label{RicciMt2M4IIA1}
 \begin{aligned}
  & \cM^t_2: \qquad R_{ab} = -  \frac{3}{k} \, \frac{1 + \l^2}{1 - \l^2} \tilde{\eta}_{ab} \, , \qquad\,\, a, b = 0 , 1 \, ,
  \\[5pt]
  & \cM_4: \qquad R_{ab} = 0 \, , \qquad\qquad\qquad\qquad a, b = 2 , 3 , 4 , 5 \, .
 \end{aligned}
\end{equation}
From the Ricci tensor it is evident that $\cM^t_2$ is an Einstein space of constant negative curvature, while $\cM_4$ is flat.

\vskip 10pt

\noindent\textbf{Comments}

\vskip 10pt

\begin{itemize}
\item The ten-dimensional line element for this solution reads
\begin{equation}
 \label{metricMt2M4IIA}
 ds^2 = ds^2_{\cM^t_2} + ds^2_{\cM_4} + \big( \fe^1 \big)^2 + \big( \fe^2 \big)^2 + \big( \fe^3 \big)^2 + \big( \fe^4 \big)^2 \, ,
\end{equation}
where again $\fe^a \, (a = 1 , \ldots , 4)$ is the frame given in \eqref{CS4lambdaFrame}. The line elements $ds^2_{\cM^t_2}$ and $ds^2_{\cM_4} $ are spanned by the frames $(e^0 , e^1)$ and $(e^2 , \ldots , e^5)$ respectively. Moreover, the normalisation of the spaces $\cM^t_2$ and $\cM_4$ follows \eqref{RicciMt2M4IIA1}.

An obvious choice for $\cM^t_2$ is $AdS_2$, while for $\cM_4$, the second line of \eqref{RicciMt2M4IIA1} suggests that we can take it to be the four-dimensional torus $T^4$.

\item On top of the metric, the NS sector of this solution contains a dilaton $\Phi$ given by \eqref{CS4lambdaScalar}. The RR sector consists of the two- and four-forms of \eqref{AnsatzM2tM4IIA}, where the parameters $c_i \, (i = 1 , \ldots , 4)$ are given in \eqref{solutionM2tM4IIA1}. The deformation parameter $\l$ in this case can take any value in $[0 , 1)$.
\end{itemize}

\subsubsection{Case $\mathbf{2}$: $\mathbf{c_2 = c_3 = 0}$}
\label{Mt4T4IIA2}

Setting $c_2 = c_3 = 0$ ensures that \eqref{paramsM2tM4IIA1} is trivially satisfied. The other two equations, \eqref{paramsM2tM4IIA2} and \eqref{paramsM2tM4IIA3}, become
\begin{equation}
 6 \, c^2_1 - 2 \, c^2_4 = \frac{6}{k} \, \frac{1 + \l^2}{1 - \l^2} \, , \qquad c^2_1 + c^2_4 = \frac{3}{k} \frac{\l}{1 - \l^2} \, .
\end{equation}
The solution for $c_1$ and $c_4$ is
\begin{equation}
 \label{solutionM2tM4IIA2}
 c^2_1 = \frac{3}{4 \, k} \frac{1 + \l + \l^2}{1 - \l^2} \, , \qquad c^2_4 = - \frac{3}{4 \, k} \frac{1 - 3 \, \l + \l^2}{1 - \l^2} \, .
\end{equation}
The requirement $c^2_1 , \, c^2_4 \geq 0$ suggests that $\l$ must lie in the interval
\begin{equation}
 \label{lambdaBoundM2tM4IIA}
 \frac{3 - \sqrt{5}}{2} \leq \l < 1 \, .
\end{equation}
The Ricci tensor on the subspaces $\cM^t_2$ and $\cM_4$ is respectively
\begin{equation}
 \label{RicciMt2M4IIA2}
 \begin{aligned}
  & \cM^t_2: \qquad R_{ab} = - \frac{3}{k} \frac{\l}{1 - \l^2} \tilde{\eta}_{ab} \, , \,\,\,\,\,\,\,\,\qquad\qquad a, b = 0 , 1 \, ,
  \\[5pt]
  & \cM_4: \qquad R_{ab} = - \frac{3}{2 \, k} \frac{1 - \l + \l^2}{1 - \l^2} \d_{ab} \, , \qquad\quad a, b = 2 , 3 , 4 , 5 \, .
 \end{aligned}
\end{equation}
Obviously, both $\cM^t_2$ and $\cM_4$ are Einstein spaces of constant negative curvature.

\vskip 10pt

\noindent\textbf{Comments}

\vskip 10pt

\begin{itemize}
 \item Here again, the line element of the ten-dimensional geometry takes the same form as \eqref{metricMt2M4IIA}. The difference now is that the spaces $\cM^t_2$ and $\cM_4$ are normalised according to \eqref{RicciMt2M4IIA2}. Therefore, both of them have now negative curvature. Hence, $\cM^t_2$ can be chosen to be $AdS_2$. On the other hand, some valid possibilities for $\cM_4$ include the four-dimensional hyperbolic space $H_4$, or the direct product of two hyperbolic planes $H_2 \times \tilde{H}_2$.
 
 \item The aforementioned geometry must be completed with the dilaton $\Phi$ given by \eqref{CS4lambdaScalar} and the RR fields in \eqref{AnsatzM2tM4IIA}. The only non-trivial parameters in this case are $c_1$, $c_4$ and they are related to the deformation $\l$ according to \eqref{solutionM2tM4IIA2}. The deformation is now restricted to take values in the interval \eqref{lambdaBoundM2tM4IIA}.
\end{itemize}

%%%%%%%%%%%%%%%%%%%%%%%%%%%%%%%%%%%%%%%%%%%%%%%%%%%%%%%%%%%%%%
\subsection{Type-IIA backgrounds on $\cM^t_4 \times \cM_2 \times CS^4_{\l}$}

We can also construct type-IIA supergravity backgrounds via a double analytic continuation to the setup of the previous section. As a result, the six dimensional space $\cM_6$ now splits as $\cM^t_4 \times \cM_2$, i.e. the time direction sits at the four-dimensional subspace $\cM^t_4$. Alternatively, one could adopt the ansatz
\begin{equation}
 \label{AnsatzM4tM2IIA}
 \begin{aligned}
  & F_0 = 0 \, ,
  \\[5pt]
  & F_2 = 2 \, e^{- \Phi} \big( c_1 \, \fe^1 \wedge \fe^3 + c_2 \, \fe^2 \wedge \fe^4 \big) \, ,
  \\[5pt]
  & F_4 = 2 \, e^{- \Phi} e^4 \wedge e^5 \wedge \big( c_3 \, \fe^1 \wedge \fe^3 + c_4 \, \fe^2 \wedge \fe^4 \big) \, ,
 \end{aligned}
\end{equation}
where $c_i \, (i = 1 , \ldots , 4)$ are constants. The above ansatz is similar to \eqref{AnsatzM2tM4IIA}, where now insted of $e^0 \wedge e^1$ in the four-form we have $e^4 \wedge e^5$. The Bianchi and flux equations \eqref{BianchisIIA}, \eqref{FluxesIIA} then imply
\begin{equation}
 \label{FormsCondsIIAM6analCont}
 d \big( e^0 \wedge e^1 \wedge e^2 \wedge e^3 \big) = d \big( e^4 \wedge e^5 \big) = 0
\end{equation}
as well as \eqref{paramsM2tM4IIA1}.

The rest of the calculation goes along the same lines as in the previous section. Nonetheless, the expressions involving the squares of the parameters $c_i \, (i = 1 , \ldots , 4)$ differ by an overall sign in $c^2_3$ and $c^2_4$. Therefore, on top of \eqref{paramsM2tM4IIA1} we have the additional conditions
\begin{equation}
 \label{paramsM4tM2IIA}
  6 \, c^2_1 + 6 \, c^2_2 + 2 \, c^2_3 + 2 \, c^2_4 = \frac{6}{k} \, \frac{1 + \l^2}{1 - \l^2} \, , \qquad c^2_1 - c^2_2 + c^2_3 - c^2_4 = \frac{3}{k} \frac{\l}{1 - \l^2} \, .
\end{equation}
These are three independent equations to be solved for a total of four parameters, i.e. $c_i \, (i = 1 , \ldots , 4)$. In turn, one of the parameters can be treated as arbitrary. Below, instead of solving \eqref{paramsM2tM4IIA1} and \eqref{paramsM4tM2IIA} in full generality, we will focus on specific cases.

Before presenting the solution of the previously mentioned algebraic constraints, it is useful to notice that the Ricci tensor on $\cM_6$ now reads
\begin{equation}
 \label{RddM6IIAanalCont}
 \begin{aligned}
  & \cM^t_4: \qquad R_{ab} = - \big( c^2_1 + c^2_2 + c^2_3 + c^2_4 \big) \hat{\eta}_{ab} \, , \qquad a, b = 0 , 1 , 2 , 3 \, ,
  \\[5pt]
  & \cM_2: \qquad R_{ab} = \big( c^2_3 + c^2_4 - c^2_1 - c^2_2 \big) \d_{ab} \, , \qquad\quad a, b = 4 , 5 \, .
 \end{aligned}
\end{equation}
Clearly, this suggests that $\cM_6$ can be expressed as the direct product $\cM^t_4 \times \cM_2$, in agreement with \eqref{FormsCondsIIAM6analCont}.

\subsubsection{Case $\mathbf{1}$: $\mathbf{c_1 = c_2 = 0}$}
\label{Mt4M2IIA1}

Setting $c_1 = c_2 = 0$ amounts to turning off the two-form $F_2$. In this case \eqref{paramsM4tM2IIA} reduce to
\begin{equation}
 c^2_3 + c^2_4 = \frac{3}{k} \, \frac{1 + \l^2}{1 - \l^2} \, , \qquad\qquad c^2_3 - c^2_4 = \frac{3}{k} \frac{\l}{1 - \l^2} \, .
\end{equation}
The solution for $c_3$ and $c_4$ is
\begin{equation}
 \label{solutionM4tM2IIA1}
 c^2_3 = \frac{3}{2 \, k} \, \frac{1 + \l + \l^2}{1 - \l^2} \, , \qquad\qquad c^2_4 = \frac{3}{2 \, k} \, \frac{1 - \l + \l^2}{1 - \l^2}
\end{equation}
%$
and $c^2_3 \, , c^2_4 \geq 0$ for all values of $\l$ in $[0 , 1)$. The Ricci tensor on $\cM^t_4$ and $\cM_2$ in this case is
\begin{equation}
 \label{RicciMt4M2IIA1}
 \begin{aligned}
  & \cM^t_4: \qquad R_{ab} = - \big( c^2_3 + c^2_4 \big) \hat{\eta}_{ab} = - \frac{3}{k} \, \frac{1 + \l^2}{1 - \l^2} \hat{\eta}_{ab} \, , \qquad\, a, b = 0 , 1 , 2 , 3 \, ,
  \\[5pt]
  & \cM_2: \qquad R_{ab} = \big( c^2_3 + c^2_4 \big) \d_{ab} = \frac{3}{k} \, \frac{1 + \l^2}{1 - \l^2} \d_{ab} \, , \qquad\qquad a, b = 4 , 5 \, .
 \end{aligned}
\end{equation}
From the above it is obvious that both $\cM^t_4$ and $\cM_2$ are Einstein spaces, where the first has negative curvature and the second positive.

\vskip 10pt

\noindent\textbf{Comments}

\vskip 10pt

\begin{itemize}
 \item The geometry of the ten-dimensional background is described by the line element
 \begin{equation}
  \label{metricMt4M2IIA}
  ds^2 = ds^2_{\cM^t_4} + ds^2_{\cM_2} + \big( \fe^1 \big)^2 + \big( \fe^2 \big)^2 + \big( \fe^3 \big)^2 + \big( \fe^4 \big)^2 \, ,
 \end{equation}
 where the frame $\fe^a \, (a = 1 , \ldots , 4)$ is by \eqref{CS4lambdaFrame}. Here $ds^2_{\cM^t_4}$ and $ds^2_{\cM_2}$ are the line elements on $\cM^t_4$ and $\cM_2$ spanned by the frames $(e^0 , e^1 , e^2 , e^3)$ and $(e^4 , e^5)$ respectively. The normalisation of $\cM^t_4$ and $\cM_2$ is dictated by \eqref{RicciMt4M2IIA1}.
 
 From \eqref{RicciMt4M2IIA1} we see that it is consistent to choose $\cM^t_4$ to be $AdS_4$ or the direct product of an $AdS_2$ and a hyperbolic plane, i.e. $AdS_2 \times H_2$. On the other hand, $\cM_2$ can be taken to be a two-sphere $S^2$.
 
 \item Aside to the aforementioned geometry, the NS sector also includes a dilaton $\Phi$ given by \eqref{CS4lambdaScalar}. The RR fields for this background can be found in \eqref{AnsatzM4tM2IIA}, with $c_1 = c_2 = 0$ and $c_3$, $c_4$ being \eqref{solutionM4tM2IIA1}. The parameter $\l$ takes values in $[0 , 1)$. 
\end{itemize}

\subsubsection{Case $\mathbf{2}$: $\mathbf{c_1 = c_4 = 0}$}
\label{Mt4M2IIA2}

The second possibility we would like to take into account is when $c_1 = c_4 = 0$. In this case \eqref{paramsM4tM2IIA} become
\begin{equation}
 3 \, c^2_2 + c^2_3 = \frac{3}{k} \, \frac{1 + \l^2}{1 - \l^2} \, , \qquad\qquad c^2_3 - c^2_2 = \frac{3}{k} \frac{\l}{1 - \l^2} \, .
\end{equation}
Solving the above we obtain $c_2$ and $c_3$
\begin{equation}
 \label{solutionM4tM2IIA2}
 c^2_2 = \frac{3}{4 \, k} \, \frac{1 - \l + \l^2}{1 - \l^2} \, , \qquad\qquad c^2_3 = \frac{3}{4 \, k} \, \frac{1 + 3 \, \l + \l^2}{1 - \l^2} \, .
\end{equation}
Obviously, $c^2_2 \, , c^2_3 \geq 0$ for all values of $\l$ in $[0 , 1)$. The Ricci tensor on $\cM^t_4$ and $\cM_2$ in this case is
\begin{equation}
 \label{RicciMt4M2IIA2}
 \begin{aligned}
  & \cM^t_4: \qquad R_{ab} = - \big( c^2_2 + c^2_3 \big) \hat{\eta}_{ab} = - \frac{3}{2 \, k} \, \frac{1 + \l + \l^2}{1 - \l^2} \hat{\eta}_{ab} \, , \qquad\,\,\, a, b = 0 , 1 , 2 , 3 \, ,
  \\[5pt]
  & \cM_2: \qquad R_{ab} = \big( c^2_3 - c^2_2 \big) \d_{ab} = \frac{3}{k} \frac{\l}{1 - \l^2} \d_{ab} \, , \qquad\qquad\qquad\quad a, b = 4 , 5 \, .
 \end{aligned}
\end{equation}
As can be inferred from the components of the Ricci tensor, $\cM^t_4$ is an Einstein space of negative constant curvature, while $\cM_2$ is an Einstein space of positive constant curvature. When $\l = 0$, the space $\cM_2$ becomes flat and it can be safely dentified with a two-dimensional torus $T^2$.

\vskip 10pt

\noindent\textbf{Comments}

\vskip 10pt

\begin{itemize}
 \item The geometry of this background is still expressed in terms of a metric that has the form \eqref{metricMt4M2IIA}, whereas now the spaces $\cM^t_4$ and $\cM_2$ are normalised according to \eqref{RicciMt4M2IIA2}. It follows from \eqref{RicciMt4M2IIA2} that we can again select $\cM^t_4$ to be either $AdS_4$ or $AdS_2 \times H_2$, and $\cM_2$ to be $S^2$.
 
 \item The other fields of the supergravity solution are the dilaton $\Phi$, given by \eqref{CS4lambdaScalar}, and the RR forms of \eqref{AnsatzM4tM2IIA}. In this case, $c_1 = c_4 = 0$, and the parameters $c_2$, $c_3$ are \eqref{solutionM4tM2IIA2}. The parameter $\l$ still takes values in $[0 , 1)$. 
\end{itemize}

\subsubsection{Case $\mathbf{3}$: $\mathbf{c_2 = c_3 = 0}$}
\label{Mt4M2IIA3}

The last option we would like to consider is when $c_2 = c_3 = 0$. Now the conditions \eqref{paramsM4tM2IIA} become
\begin{equation}
 3 \, c^2_1 + c^2_4 = \frac{3}{k} \, \frac{1 + \l^2}{1 - \l^2} \, , \qquad c^2_1 - c^2_4 = \frac{3}{k} \frac{\l}{1 - \l^2} \, .
\end{equation}
The solution for $c_1$ and $c_4$ is
\begin{equation}
 \label{solutionM4tM2IIA3}
 c^2_1 = \frac{3}{4 \, k} \, \frac{1 + \l + \l^2}{1 - \l^2} \, , \qquad\qquad c^2_4 = \frac{3}{4 \, k} \, \frac{1 - 3 \, \l + \l^2}{1 - \l^2} \, .
\end{equation}
Requiring $c^2_1 \, , c^2_4 \geq 0$ impose the bound \eqref{lambdaBoundM6tIIA} on $\l$.

For the Ricci tensor on $\cM^t_4$ and $\cM_2$ we find
\begin{equation}
 \label{RicciMt4M2IIA3}
 \begin{aligned}
  & \cM^t_4: \qquad R_{ab} = - \big( c^2_1 + c^2_4 \big) \hat{\eta}_{ab} = - \frac{3}{2 \, k} \, \frac{1 - \l + \l^2}{1 - \l^2} \hat{\eta}_{ab} \, , \qquad\quad\,\, a, b = 0 , 1 , 2 , 3 \, ,
  \\[5pt]
  & \cM_2: \qquad R_{ab} = \big( c^2_4 - c^2_1 \big) \d_{ab} = - \frac{3}{k} \frac{\l}{1 - \l^2} \d_{ab} \, , \qquad\qquad\qquad\quad a, b = 4 , 5 \, .
 \end{aligned}
\end{equation}
Clearly, both $\cM^t_4$ and $\cM_2$ are Einstein spaces of constant negative curvature. The space $\cM_2$ becomes flat when $\l = 0$; in this case, it can be taken to be a two-dimensional torus $T^2$.

\vskip 10pt

\noindent\textbf{Comments}

\vskip 10pt

\begin{itemize}
 \item As in the previous two examples, the geometry takes the form \eqref{metricMt4M2IIA}. The difference now is that both $\cM^t_4$ and $\cM_2$ have negative curvature as is understood from \eqref{RicciMt4M2IIA3}. Therefore in this case, $\cM^t_4$ can still be taken to be either $AdS_4$ or $AdS_2 \times H_2$, but $\cM_2$ is identified with a hyperbolic plane $H_2$.
 
 \item Along with the metric, the supergravity solution also comes with a dilaton $\Phi$, given by \eqref{CS4lambdaScalar}, and the RR fields of \eqref{AnsatzM4tM2IIA}. In this case, $c_2 = c_3 = 0$, while for $c_1$ and $c_4$ we find \eqref{solutionM4tM2IIA3}, with $\l$ taking values in \eqref{lambdaBoundM6tIIA}. 
\end{itemize}

%%%%%%%%%%%%%%%%%%%%%%%%%%%%%%%%%%%%%%%%%%%%%%%%%%%%%%%%%%%%%%
\subsection{Type-IIB backgrounds on $\cM^t_2 \times \cM_3 \times S^1 \times CS^4_{\l}$}
\label{Mt2T4IIB}

We now turn to the construction of solutions of type-IIB supergravity, accommodating the $CS^4_{\l}$ model. With this in mind, we propose the following ansatz for the type-IIB RR fields
\begin{equation}
 \label{AnsatzM2tM3S1IIB}
 \begin{aligned}
  & F_1 = 0 \, ,
  \\[5pt]
  & F_3 = 2 \, e^{- \Phi} e^5 \wedge \big( c_1 \, \fe^1 \wedge \fe^3 + c_2 \, \fe^2 \wedge \fe^4 \big) \, ,
  \\[5pt]
  & F_5 = 2 \, e^{- \Phi} \big( 1 + \star \big) e^2 \wedge e^3 \wedge e^4 \wedge \big( c_3 \, \fe^1 \wedge \fe^3 + c_4 \, \fe^2 \wedge \fe^4 \big) \, ,
 \end{aligned}
\end{equation}
where as usual, $c_i \, (i = 1 , \ldots , 4)$ are constants to be fixed. The Bianchi and flux equations \eqref{BianchisIIB}, \eqref{FluxesIIB} now imply
\begin{equation}
 \label{FormsCondsIIAM1tM3S1}
 d e^5 = d \big( e^0 \wedge e^1 \big) = d \big( e^2 \wedge e^3 \wedge e^4 \big) = 0 \, .
\end{equation}
and an algebraic constraint on the parameters $c_i$ which takes the form
\begin{equation}
 \label{paramsM2tM3S1IIB1}
 c_1 \, c_4 + c_2 \, c_3 = 0 \, .
\end{equation}
The conditions \eqref{FormsCondsIIAM1tM3S1} suggest that the six-dimensional space $\cM_6$, transverse to the geometry of $CS^4_{\l}$, takes the direct product form $\cM^t_2 \times \cM_3 \times S^1$.

Like in all previous cases, the Einstein equations \eqref{EinsteinIIB} can tell us more about the geometry $\cM_6$. Starting with the non-vanishing components of $\cT^{IIB}_{ab}$ in the frame $e^a \, (a = 0 , 1 , \ldots , 9)$ we find
\begin{equation}
 \label{TddM2tM3S1IIB}
 \begin{aligned}
  & \cT^{IIB}_{ab} = - \big( c^2_1 + c^2_2 + c^2_3 + c^2_4 \big) \tilde{\eta}_{ab} \, , \qquad\, a, b = 0 , 1 \, ,
  \\[5pt]
  & \cT^{IIB}_{ab} = \big( c^2_3 + c^2_4 - c^2_1 - c^2_2 \big) \d_{ab} \, , \qquad\quad a, b = 2 , 3 , 4 \, ,
  \\[5pt]
  & \cT^{IIB}_{55} = \big( c^2_1 + c^2_2 - c^2_3 - c^2_4 \big) \, ,
  \\[5pt]
  & \cT^{IIB}_{ab} = \big( c^2_1 - c^2_2 + c^2_3 - c^2_4 \big) \bar{\eta}_{ab} \, , \qquad\quad a, b = 6 , 7 , 8 , 9 \, .
 \end{aligned}
\end{equation}
The first three lines above combined with \eqref{EinsteinIIB} imply that the non-zero components of the Ricci tensor on $\cM_6$ are
\begin{equation}
 \label{RicciM2tM3S1IIB}
 \begin{aligned}
  \cM^t_2:& \qquad R_{ab} = - \big( c^2_1 + c^2_2 + c^2_3 + c^2_4 \big) \tilde{\eta}_{ab} \, , \qquad\, a, b = 0 , 1 \, ,
  \\[5pt]
  \cM_3:& \qquad R_{ab} = \big( c^2_3 + c^2_4 - c^2_1 - c^2_2 \big) \d_{ab} \, , \qquad\quad a, b = 2 , 3 , 4 \, ,
  \\[5pt]
  S^1:& \qquad R_{55} = \big( c^2_1 + c^2_2 - c^2_3 - c^2_4 \big) \, .
 \end{aligned}
\end{equation}
Knowing the Ricci tensor on $\cM_6$ we can compute the Ricci scalar $R_{\cM_6}$, and after combining it with \eqref{RicciScalarM6} we find
\begin{equation}
 \label{paramsM2tM3S1IIB2}
 c^2_1 + c^2_2 = \frac{3}{2 \, k} \, \frac{1 + \l^2}{1 - \l^2} \, .
\end{equation}
Moreover, since $e^5$ corresponds to the direction of the circle $S^1$, we should require $R_{55} = 0$, which translates to
\begin{equation}
 \label{paramsM2tM3S1IIB3}
 c^2_1 + c^2_2 - c^2_3 - c^2_4 = 0 \, .
\end{equation}
A fourth condition on the constants $c_i$ can be obtained from the last line of \eqref{TddM2tM3S1IIB}, in view of the Einstein equations \eqref{EinsteinIIB} and the second property in \eqref{CS4lambdaBetaFunctions}
\begin{equation}
 \label{paramsM2tM3S1IIB4}
 c^2_1 - c^2_2 + c^2_3 - c^2_4 = \frac{3}{k} \frac{\l}{1 - \l^2} \, .
\end{equation}
In total, we have to solve four algebraic equations -- \eqref{paramsM2tM3S1IIB1}, \eqref{paramsM2tM3S1IIB2}, \eqref{paramsM2tM3S1IIB3} and \eqref{paramsM2tM3S1IIB4} -- for the set of four constants $c_i \, (i = 1 , \ldots , 4)$. The solution is given in terms of the constants $k$ and $\l$ through the combinations
\begin{equation}
 \label{solutionM2tM3S1IIB}
 c_i = \frac{s_i}{2} \sqrt{\frac{3}{k} \frac{1 + \l + \l^2}{1 - \l^2}} \qquad (i = 1 , 3) \, , \qquad c_i = \frac{s_i}{2} \sqrt{\frac{3}{k} \frac{1 - \l + \l^2}{1 - \l^2}} \qquad (i = 2 , 4) \, ,
\end{equation}
where $s^2_i = 1 \, (i = 1 , \ldots , 4)$. The constants $s_i$ satisfy the condition
\begin{equation}
 \label{solutionM2tM3S1IIBsConstraint}
 s_1 \, s_4 + s_2 \, s_3 = 0 \, ,
\end{equation}
which reflects \eqref{paramsM2tM3S1IIB1}. From \eqref{solutionM2tM3S1IIB} is evident that all parameters $c_i \, (i = 1 , \ldots , 4)$ are real for $\l \in [0 , 1)$.

The Ricci tensor components on $\cM_6$ for the solution \eqref{solutionM2tM3S1IIB} read
\begin{equation}
 \label{RicciM2tT3S1IIB}
 \begin{aligned}
  \cM^t_2:& \qquad R_{ab} = - \frac{3}{k} \, \frac{1 + \l^2}{1 - \l^2} \tilde{\eta}_{ab} \, , \qquad a, b = 0 , 1 \, ,
  \\[5pt]
  \cM_3:& \qquad R_{ab} = 0 \, , \qquad\qquad\qquad\quad\,\, a, b = 2 , 3 , 4 \, ,
  \\[5pt]
  S^1:& \qquad R_{55} = 0 \, .
 \end{aligned}
\end{equation}
As a result, the space $\cM_3$ is flat, and thus its combination with the direction $e^5$ on the circle $S^1$ can be viewed as a four-dimensional torus $T^4$.

\vskip 10pt

\noindent\textbf{Comments}

\vskip 10pt

\begin{itemize}
 \item The metric of the type-IIB solution discussed in this section reads
 \begin{equation}
  ds^2 = ds^2_{\cM^t_2} + ds^2_{T^4} + \big( \fe^1 \big)^2 + \big( \fe^2 \big)^2 + \big( \fe^3 \big)^2 + \big( \fe^4 \big)^2 \, ,
 \end{equation}
 where $\fe^a \, (a = 1 , \ldots , 4)$ stands for the frame \eqref{CS4lambdaFrame} on the target space of $CS^4_{\l}$. Also, $\cM^t_2$ is spanned by $(e^0 , e^1)$ and is normalised according to \eqref{RicciM2tT3S1IIB}, while the four-torus $T^4$ is spanned by $(e^2 , e^3 , e^4 , e^5)$. The two-dimensional space $\cM^t_2$ can then be chosen as $AdS_2$.
 
 \item The other fields that contribute to the supergravity solution are the dilaton $\Phi$ given in \eqref{CS4lambdaScalar}, and the RR forms of \eqref{AnsatzM2tM3S1IIB}. The constants $c_i \, (i = 1 , \ldots , 4)$ that enter the ansatz \eqref{AnsatzM2tM3S1IIB} satisfy \eqref{solutionM2tM3S1IIB}, \eqref{solutionM2tM3S1IIBsConstraint} and $\l$ takes values in $[0 , 1)$.
 
 \item The backgrounds discussed here and in Section \ref{Mt4T4IIA1} are related via T-duality in the torus direction $e^5$.
 
\end{itemize}

%%%%%%%%%%%%%%%%%%%%%%%%%%%%%%%%%%%%%%%%%%%%%%%%%%%%%%%%%%%%%%
\subsection{Comments on backgrounds with $CH_{4, \l}$}

In this section, we discuss how to promote the $CH_{4, \l}$ model into a full solution of the type-II supergravity. To this end, one can follow the same approach as for the $CS^4_{\l}$. That is, to complete the four-dimensional target space of $CH_{4, \l}$, described in Sec. \ref{CS4Lambda}, with a six-dimensional manifold $\cM_6$ and propose appropriate ansatze for the RR sector to gain further insight for the structure of $\cM_6$. Equivalently, these solutions can be obtained from those for $CS^4_{\l}$, found in the previous sections, by replacing the background fields of $CS^4_{\l}$ given in \eqref{CS4lambdaFrame}, \eqref{CS4lambdaScalar}, with those of $CH_{4, \l}$ in \eqref{CH4lambdaFrame}, \eqref{CH4lambdaScalar}. Effectively, this amounts to replacing $k$ by $-k$. However, imposing reality of the background eliminates some of the possibilities. This leaves only the construction in Sec. \ref{Mt2M4CS4lambdaIIA} as a valid starting point.

In more detail, we assume that the ten-dimensional spacetime decomposes as $\cM_6 \times CH_{4, \l}$, where the metric on $\cM_6$ is a expressed by the frame $(e^0 , \ldots , e^5)$ as in \eqref{MetricM6}. For the remaining four directions we identify $e^6 \to \hfe^1$, $e^7 \to \hfe^2$, $e^8 \to \hfe^3$ and $e^9 \to \hfe^4$, with $\hfe^a \, (a = 1 , \ldots , 4)$ being \eqref{CH4lambdaFrame}. The difference compared to the solutions on $\cM_6 \times CS^4_{\l}$ now is that the Ricci scalar on $\cM_6$ is given by
\begin{equation}
 \label{RicciScalarM6CH4lambda}
 R_{\cM_6} - \frac{6}{k} \, \frac{1 + \l^2}{1 - \l^2} = 0 \, .
\end{equation}
Additionally, we take the NS three-form $H$ to be zero and the dilaton $\Phi$ to be given by \eqref{CH4lambdaScalar}.

Following Sec. \ref{Mt2M4CS4lambdaIIA}, the analogue of \eqref{AnsatzM2tM4IIA} is
\begin{equation}
 \label{AnsatzM2tM4CH4lambdaIIA}
 \begin{aligned}
  & F_0 = 0 \, ,
  \\[5pt]
  & F_2 = 2 \, e^{- \Phi} \big( c_1 \, \hfe^1 \wedge \hfe^3 + c_2 \, \hfe^2 \wedge \hfe^4 \big) \, ,
  \\[5pt]
  & F_4 = 2 \, e^{- \Phi} e^0 \wedge e^1 \wedge \big( c_3 \, \hfe^1 \wedge \hfe^3 + c_4 \, \hfe^2 \wedge \hfe^4 \big) \, ,
 \end{aligned}
\end{equation}
with $c_i \, (i = 1 , \ldots , 4)$ being constants. Consistency of the ansatz with the Bianchi and flux equations \eqref{BianchisIIA} and \eqref{FluxesIIA} implies \eqref{FormsCondsIIAM6} and \eqref{paramsM2tM4IIA1}.

Focusing on the Einstein equations \eqref{EinsteinIIA}, we find that the non-vanishing components of $ \cT^{IIA}_{ab}$ in the frame $e^a \, (a = 0 , 1 , \ldots , 9)$ are again given by \eqref{TddM6IIA}. As a result, the six-dimensional space $\cM_6$ again splits as $\cM^t_2 \times \cM_4$, with $\cM^t_2$ spanned by $(e^0 , e^1)$ and $\cM_4$ by $(e^2 , e^3 , e^4 , e^5)$. The corresponding Ricci tensors are also given by \eqref{RddM6IIA}. The rest of the computation goes through as in Sec. \ref{Mt2M4CS4lambdaIIA}. The difference now is that the constraints \eqref{paramsM2tM4IIA2} and \eqref{paramsM2tM4IIA3} are mapped to
\begin{equation}
 \label{paramsM2tM4CH4lambdaIIA23}
 \begin{aligned}
  & 2 \, c^2_3 + 2 \, c^2_4 - 6 \, c^2_1 - 6 \, c^2_2 = \frac{6}{k} \, \frac{1 + \l^2}{1 - \l^2} \, ,
  \\[5pt]
  & c^2_2 - c^2_1 + c^2_3 - c^2_4 = \frac{3}{k} \frac{\l}{1 - \l^2} \, .
 \end{aligned}
\end{equation}
The above, together with \eqref{paramsM2tM4IIA1} consist of three independent algebraic constraints for the four variables $c_i \, ( i = 1 , \ldots , 4)$. Therefore, one of the parameters $c_i$ can be treated as arbitrary. Note that all three equations, \eqref{paramsM2tM4IIA1} and \eqref{paramsM2tM4CH4lambdaIIA23} satisfy the symmetry \eqref{Symmetryc1c2c3c4}, and thus different solutions must be related by this symmetry. Nevertheless, here we will not attempt to solve the system \eqref{paramsM2tM4IIA1} and \eqref{paramsM2tM4CH4lambdaIIA23} in its full generality, but instead we will focus on a specific case. In particular, we will consider $c_1 = c_2 = 0$. We could also take $c_1 = c_4 = 0$ or $c_2 = c_3 = 0$, nevertheless these would correspond to complex solutions for the remaining parameters $c_i$.

\subsubsection{Case: $\mathbf{c_1 = c_2 = 0}$}
\label{M2tM4CH4lambdaIIA}

In this case the RR sector only contains a four-form $F_4$. The condition \eqref{paramsM2tM4IIA1} is automatically satisfied, while \eqref{paramsM2tM4CH4lambdaIIA23} become
\begin{equation}
 c^2_3 + c^2_4 = \frac{3}{k} \, \frac{1 + \l^2}{1 - \l^2} \, , \qquad c^2_3 - c^2_4 = \frac{3}{k} \frac{\l}{1 - \l^2} \, .
\end{equation}
Solving for $c_3$ and $c_4$ we find
\begin{equation}
 \label{solutionM2tM4CH4lambdaIIA1}
 c^2_3 = \frac{3}{2 \, k} \, \frac{1 + \l + \l^2}{1 - \l^2} \, , \qquad\qquad c^2_4 = \frac{3}{2 \, k} \, \frac{1 - \l + \l^2}{1 - \l^2} \, .
\end{equation}
From the last it is evident that $c^2_3 , \, c^2_4 \geq 0$ for all values of $\l$ in $[0 , 1)$.

The Ricci tensors on $\cM^t_2$ and $\cM_4$ for the above values of the parameters $c_i \, (i = 1 , \ldots , 4)$ read
\begin{equation}
 \label{RicciMt4M2CH4lambdaIIA1}
 \begin{aligned}
  & \cM^t_2: \qquad R_{ab} = - \frac{3}{k} \, \frac{1 + \l^2}{1 - \l^2} \tilde{\eta}_{ab} \, , \qquad a, b = 0 , 1 \, ,
  \\[5pt]
  & \cM_4: \qquad R_{ab} = \frac{3}{k} \, \frac{1 + \l^2}{1 - \l^2} \d_{ab} \, , \qquad\quad a, b = 2 , 3 , 4 , 5 \, .
 \end{aligned}
\end{equation}
As a result, $\cM^t_2$ and $\cM_4$ are Einstein spaces, where the first one has negative constant curvature, while the curvature of the second is positive.

\vskip 10pt

\noindent\textbf{Comments}

\vskip 10pt

\begin{itemize}
 \item The ten-dimensional geometry of the supergravity background is now described by the line element
 \begin{equation}
  \label{metricMt2M4CH4lambdaIIA}
  ds^2 = ds^2_{\cM^t_2} + ds^2_{\cM_4} + \big( \hfe^1 \big)^2 + \big( \hfe^2 \big)^2 + \big( \hfe^3 \big)^2 + \big( \hfe^4 \big)^2 \, .
 \end{equation}
 The frame components $\hfe^a \, (a = 1 , \ldots , 4)$ are now given by \eqref{CH4lambdaFrame}. The line elements $ds^2_{\cM^t_2}$ and $ds^2_{\cM_4}$ for the spaces $\cM^t_2$ and $\cM_4$ are spanned by the frames $(e^0 , e^1)$ and $(e^2 , e^3 , e^4 , e^5)$ respectively. Moreover, $\cM^t_2$ and $\cM_4$ are normalised according to \eqref{RicciMt4M2CH4lambdaIIA1}.
 
 Additionally, from \eqref{RicciMt4M2CH4lambdaIIA1}, it follows that $\cM^t_2$ can be chosen as $AdS_2$. Meanwhile, $\cM_4$ can be taken as a four-dimensional sphere $S^4$, the complex projective space $\mathbb{CP}^2$, or the direct product $S^2 \times \tilde{S}^2$ of two two-dimensional spheres.
 
 \item The NS sector of the supergravity background discussed here also includes a dilaton $\Phi$ given by \eqref{CH4lambdaScalar}, while the NS three-form vanishes. For the RR fields we find that the two-form $F_2$ is trivial, while the four-form $F_4$ is given by \eqref{AnsatzM2tM4CH4lambdaIIA}, with $c_3$, $c_4$ being \eqref{solutionM2tM4CH4lambdaIIA1}. The supergravity solution now makes sense for all values of $\l$ in $[0 , 1)$.
\end{itemize}

%%%%%%%%%%%%%%%%%%%%%%%%%%%%%%%%%%%%%%%%%%%%%%%%%%%%%%%%%%%%%%
\section{Conclusions}
\label{Section5}

In this work, we constructed various solutions of type-II supergravity by uplifting the $\l$-models on $\nicefrac{SO(4)_k}{SO(3)_k}$, $\nicefrac{SO(5)_k}{SO(4)_k}$ and the non-compact version of the latter. Our approach relies on introducing well-motivated ansatze for the RR fields, which define the structure of the corresponding geometries. The ten-dimensional manifolds are of the form $\cM_7 \times CS^3_{\l}$, $\cM_6 \times CS^4_{\l}$ and $\cM_6 \times CH_{4, \l}$, where the transverse spaces $\cM_6$ and $\cM_7$ split into direct products of Einstein spaces with constant curvature. Notably, in all examples examined here, the transverse geometries allow for the presence of AdS factors. In particular the AdS backgrounds discussed here are listed in Tables \ref{ListOfSolutionsCS3lambda} \& \ref{ListOfSolutionsCS4lambda}.
\begin{table}[h!]
\center
\setlength{\arrayrulewidth}{0.4mm}
\begin{tabular}{ |c|c|c| } 
 \hline
 \textbf{Geometry} & \textbf{Type} & \textbf{Section} \\ 
 \hline
 &	&	\\[-10pt]
 $AdS_2 \times H_2 \times S^3 \times CS^3_{\l }$ & IIA & \ref{Mt4M3CS3lambdaIIA} \\[2pt]
 $AdS_2 \times H_2 \times H_3 \times CS^3_{\l}$ & IIB & \ref{Mt4M3CS3lambdaIIB} \\[2pt] 
 $AdS_3 \times H_2 \times \tilde{H}_2 \times CS^3_{\l}$ & IIB & \ref{Mt3M4CS3lambdaIIA} \\[2pt]
 $AdS_3 \times H_4 \times CS^3_{\l}$ & IIB & \ref{Mt3M4CS3lambdaIIA} \\[2pt]
 $AdS_4 \times S^3 \times CS^3_{\l }$ & IIA & \ref{Mt4M3CS3lambdaIIA} \\[2pt]
 $AdS_4 \times H_3 \times CS^3_{\l}$ & IIB & \ref{Mt4M3CS3lambdaIIB} \\[2pt] 
 \hline
\end{tabular}
\caption{List of type-II solutions based on $CS^3_{\l}$ and where they are discussed in the main text. The notation $CS^3_{\l}$ refers to the target space of the $\l$-deformed coset $\nicefrac{SO(4)_k}{SO(3)_k}$.
\label{ListOfSolutionsCS3lambda}}
\end{table}
\begin{table}[h!]
\center
\setlength{\arrayrulewidth}{0.4mm}
\begin{tabular}{ |c|c|c| } 
 \hline
 \textbf{Geometry} & \textbf{Type} & \textbf{Section} \\ 
 \hline
 &	&	\\[-10pt]
 $AdS_2 \times H_2 \times \tilde{H}_2 \times CS^4_{\l }$ & IIA & \ref{M6CS4lambdaIIA}, \ref{Mt4T4IIA2}, \ref{Mt4M2IIA3} \\[2pt]
 $AdS_2 \times H_2 \times S^2 \times CS^4_{\l }$ & IIA & \ref{Mt4M2IIA1}, \ref{Mt4M2IIA2} \\[2pt]
 $AdS_2 \times S^2 \times \tilde{S}^2 \times CH_{4, \l}$ & IIA & \ref{M2tM4CH4lambdaIIA} \\[2pt]
 $AdS_2 \times H_4 \times CS^4_{\l }$ & IIA & \ref{M6CS4lambdaIIA}, \ref{Mt4T4IIA2} \\[2pt]
 $AdS_2 \times S^4 \times CH_{4, \l}$ & IIA & \ref{M2tM4CH4lambdaIIA} \\[2pt]
 $AdS_2 \times \mathbb{CP}^2 \times CH_{4, \l}$ & IIA & \ref{M2tM4CH4lambdaIIA} \\[2pt]
 $AdS_2 \times T^4 \times CS^4_{\l }$ & IIA & \ref{Mt4T4IIA1} \\[2pt]
 $AdS_2 \times T^4 \times CS^4_{\l }$ & IIB & \ref{Mt2T4IIB} \\[2pt]  
 $AdS_3 \times H_3 \times CS^4_{\l }$ & IIA & \ref{M6CS4lambdaIIA} \\[2pt]
 $AdS_4 \times H_2 \times CS^4_{\l }$ & IIA & \ref{M6CS4lambdaIIA}, \ref{Mt4M2IIA3} \\[2pt]
 $AdS_4 \times S^2 \times CS^4_{\l }$ & IIA & \ref{Mt4M2IIA1}, \ref{Mt4M2IIA2} \\[2pt]
 $AdS_6 \times CS^4_{\l }$ & IIA & \ref{M6CS4lambdaIIA} \\[2pt]   
 \hline
\end{tabular}
\caption{List of type-II solutions based on $CS^4_{\l}$, $CH_{4, \lambda}$ and where they are discussed in the main text. The notation $CS^4_{\l}$ and $CH_{4, \l}$ refers to the target spaces of the $\l$-deformed cosets $\nicefrac{SO(5)_k}{SO(4)_k}$ and $\nicefrac{SO(1,4)_{-k}}{SO(4)_{-k}}$ respectively.}
\label{ListOfSolutionsCS4lambda}
\end{table}
More solutions can be constructed by generalising the ansatze for the RR fields examined in the main text. For example, instead of \eqref{Mt2M4CS4lambdaIIA} one could consider
\begin{equation}
 \begin{aligned}
  F_0 & = 0 \, ,
  \\[5pt]
  F_2 & = 2 \, e^{- \Phi} \big( c_1 \, \fe^1 \wedge \fe^3 + c_2 \, \fe^2 \wedge \fe^4 \big) \, ,
  \\[5pt]
  F_4 & = 2 \, e^{- \Phi} e^0 \wedge e^1 \wedge \big( c_3 \, \fe^1 \wedge \fe^3 + c_4 \, \fe^2 \wedge \fe^4 \big)
  \\[5pt]
  & + 2 \, e^{- \Phi} e^2 \wedge e^3 \wedge \big( c_5 \, \fe^1 \wedge \fe^3 + c_6 \, \fe^2 \wedge \fe^4 \big)
  \\[5pt]
  & + 2 \, e^{- \Phi} e^4 \wedge e^5 \wedge \big( c_7 \, \fe^1 \wedge \fe^3 + c_8 \, \fe^2 \wedge \fe^4 \big) \, ,
 \end{aligned}
\end{equation}
with $c_i , \, (i = 1 , \ldots , 8)$ being constants. Similarly for \eqref{AnsatzM2tM4CH4lambdaIIA}. However, an exhaustive treatment of all possibilities is beyond the scope of this paper.

The amount of supersymmetry preserved by the supergravity backgrounds constructed in this work still needs to be investigated. Some of the AdS solutions found in the main text may appear to be non-supersymmetric, raising questions about their perturbative stability. Another question that arises naturally is whether the presence of RR fluxes affects the integrability of the corresponding full string solutions. It would also be interesting to determine if the backgrounds found here originate from brane intersections. Finally, given that all the examples we focus on contain AdS factors, one could explore these backgrounds within the framework of the AdS/CFT correspondence. This approach would bring $\l$-deformations in contact with the ideas of holography. In the non-Abelian T-dual limit, this connection opened new avenues for understanding holographic dual field theories, e.g. \cite{Sfetsos:2010uq,Itsios:2013wd,Lozano:2016kum,Lozano:2016wrs,Lozano:2017ole,Itsios:2017cew,Itsios:2017nou}.

%%%%%%%%%%%%%%%%%%%%%%%%%%%%%%%%%%%%%%%%%%%%%%%%%%%%%%%%%%%%%%
\section*{Acknowledgements}

This research is supported by the Einstein Stiftung Berlin via the Einstein International Postdoctoral Fellowship program ``Generalised dualities and their holographic applications to condensed matter physics'' (project number IPF- 2020-604). I would also like to acknowledge support by the Deutsche Forschungsgemeinschaft (DFG, German Research Foundation) via the Emmy Noether program ``Exploring the landscape of string theory flux vacua using exceptional field theory'' (project number 426510644).

\appendix

%%%%%%%%%%%%%%%%%%%%%%%%%%%%%%%%%%%%%%%%%%%%%%%%%%%%%%%%%%%%%%
\section{The type-IIA and type-IIB supergravities}
\label{AppA}

Here, we present a concise overview of the equations of motion for the type-IIA and type-IIB supergravity theories.

%%%%%%%%%%%%%%%%%%%%%%%%%%%%%%%%%%%%%%%%%%%%%%%%%%%%%%%%%%%%%

\subsection{The type-IIA supergravity}

The field content of the NS sector in type-IIA supergravity consists of the ten-dimensional metric $g_{\m\n}$, the dilaton $\Phi$ and a three-form $H$. The RR sector contains the forms $F_2$ and $F_4$, where the index indicates their rank. The form fields can be expressed in terms of the potentials $(B , C_1 , C_3)$ as
\begin{equation}
 H = dB \, , \qquad F_2 = dC_1 \, , \qquad F_4 = dC_3 - H \wedge C_1 \, .
\end{equation}
The above imply the Bianchi identities
\begin{equation}
 \label{BianchisIIA}
 dH = 0 \, , \qquad dF_2 = 0 \, , \qquad dF_4 = H \wedge F_2 \, .
\end{equation}

The dynamics of the various fields is described by the following ten-dimensional action in the string frame
\begin{equation}
\label{ActionIIA}
 \begin{aligned}
  S_{IIA} = & \frac{1}{2 \kappa^2_{10}} \int\limits_{M_{10}} \sqrt{- g} \Bigg[  e^{-2 \Phi} \left(  R + 4 \big(  \partial \Phi \big)^2 - \frac{H^2}{12}  \right) - \frac{1}{2} \left(  \frac{F^2_2}{2} + \frac{F^2_4}{4!} \right) \Bigg]
  \\[5pt]
  - & \frac{1}{4 \kappa^2_{10}} \int\limits_{M_{10}} B \wedge dC_3 \wedge dC_3 \, . 
 \end{aligned}
\end{equation}
Varying the action with respect to $\Phi$, we derive the equation of motion for the dilaton
\begin{equation}
\label{DilatonEOM}
 R + 4 \nabla^2 \Phi - 4 \big(  \partial \Phi \big)^2 - \frac{H^2}{12} = 0 \, .
\end{equation}
On the other hand, the variation with respect to the metric leads to the Einstein equations
\begin{equation}
\label{EinsteinIIA}
 \begin{aligned}
  & \cE^{IIA}_{\m\n} = R_{\m\n} + 2 \,\nabla_\m \nabla_\n \Phi - \frac{1}{4} \, \big(  H^2 \big)_{\m\n} - \cT^{IIA}_{\m\n} = 0 \, ,
  \\[5pt]
  & \cT^{IIA}_{\m\n} = e^{2 \Phi} \Bigg[  \frac{1}{2} \big(  F^2_2  \big)_{\m\n} + \frac{1}{12} \big(  F^2_4  \big)_{\m\n} - {1\ov 4}g_{\m\n} \Bigg(  \frac{F^2_2}{2} + \frac{F^2_4}{24} \Bigg)   \Bigg] \, .
 \end{aligned}
\end{equation}
Finally, the equations of motion for the potentials $(B , C_1 , C_3)$ result in the flux equations
\begin{equation}
\label{FluxesIIA}
 \begin{aligned}
  & d \big(  e^{- 2 \Phi} \star H  \big) - F_2 \wedge \star F_4 - \frac{1}{2} F_4 \wedge F_4 = 0 \, ,
  \\[5pt]
  & d \star F_2 + H \wedge \star F_4 = 0 \, ,
  \\[5pt]
  & d \star F_4 + H \wedge  F_4= 0 \, .
 \end{aligned}
\end{equation}

%%%%%%%%%%%%%%%%%%%%%%%%%%%%%%%%%%%%%%%%%%%%%%%%%%%%%%%%%%%%%

\subsection{The type-IIB supergravity}

Like in type-IIA supergravity, the NS sector of the type-IIB theory contains the ten-dimensional metric $g_{\m\n}$, the dilaton $\Phi$ and a three-form $H$. However, the RR sector now contains forms of odd rank, namely the $F_1$, $F_3$ and the self-dual five-form $F_5$. Again, the index stands for the rank of the forms. The form fields can be written in terms of the potentials $(B , C_0 , C_2 , C_4)$ as
\begin{equation}
 H = dB \, , \qquad F_1 = dC_0 \, , \qquad F_3 = dC_2 - C_0 H \, , \qquad F_5 = dC_4 - H \wedge C_2 \, .
\end{equation}
The latter imply the Bianchi identities
\begin{equation}
 \label{BianchisIIB}
 dH = 0 \, , \qquad dF_1 = 0 \, , \qquad dF_3 = H \wedge F_1 \, , \qquad dF_5 = H \wedge F_3 \, .
\end{equation}

The dynamics of this field content is now described by the following ten-dimensional action in the string frame
\begin{equation}
\label{ActionIIB}
 \begin{aligned}
  S_{IIB} = & \frac{1}{2 \kappa^2_{10}} \int\limits_{M_{10}} \sqrt{- g} \Bigg[  e^{-2 \Phi} \left(  R + 4 \big(  \partial \Phi \big)^2 - \frac{H^2}{12}  \right) - \frac{1}{2} \left(  F^2_1 + \frac{F^2_3}{3!} + \frac{F^2_5}{2 \cdot 5!} \right) \Bigg]
  \\[5pt]
  - & \frac{1}{4 \kappa^2_{10}} \int\limits_{M_{10}} C_4 \wedge H \wedge dC_2 \, . 
 \end{aligned}
\end{equation}
Since the NS sector in the type-IIB action coincides with that of the type-IIA, the equation of motion for the dilaton is still given by \eqref{DilatonEOM}. On the other hand, varying the action with respect to the metric leads to the Einstein equations
\begin{equation}
\label{EinsteinIIB}
 \begin{aligned}
  & \cE^{IIB}_{\m\n} = R_{\m\n} + 2 \,\nabla_\m \nabla_\n \Phi - \frac{1}{4} \, \big(  H^2 \big)_{\m\n} - \cT^{IIB}_{\m\n} = 0 \, ,
  \\[5pt]
  & \cT^{IIB}_{\m\n} = e^{2 \Phi} \Bigg[  \frac{1}{2} \big(  F^2_1  \big)_{\m\n} + \frac{1}{4} \big(  F^2_3  \big)_{\m\n} + \frac{1}{96} \big(  F^2_5  \big)_{\m\n} - g_{\m\n} \Bigg(  \frac{F^2_1}{4} + \frac{F^2_3}{24} \Bigg)   \Bigg] \, .
 \end{aligned}
\end{equation}
Finally, the equations of motion for the potentials $(B , C_0 , C_2 , C_4)$ are given by the flux equations
\begin{equation}
\label{FluxesIIB}
 \begin{aligned}
  & d \big(  e^{- 2 \Phi} \star H  \big) - F_1 \wedge \star F_3 - F_3 \wedge F_5 =0\, ,
  \\[5pt]
  & d \star F_1 + H \wedge \star F_3 = 0 \, ,
  \\[5pt]
  & d \star F_3 + H \wedge \star F_5 = 0
  \\[5pt]
  & d \star F_5 - H \wedge  F_3 = 0 \, .
 \end{aligned}
\end{equation}

%%%%%%%%%%%%%%%%%%%%%%%%%%%%%%%%%%%%%%%%%%%%%%%%%%%%%%%%%%%%%%%%%%%%%%%%

%\bibliographystyle{utphys}
%\bibliography{BibItsios.bib}

\end{document}